\documentclass[a4paper,12pt]{article}
\usepackage[utf8]{inputenc}

\usepackage[margin=2.5cm]{geometry}

\usepackage{amsmath}
\usepackage{amsfonts}
\usepackage{graphicx}
\usepackage{caption}
\usepackage{subcaption}
\usepackage{float}
\usepackage{comment}

\usepackage{authblk}
\usepackage{amssymb}

\usepackage{epstopdf}
\usepackage{epsfig}
\usepackage{hyperref}

\usepackage{color}
\usepackage[normalem]{ulem}
\usepackage{xcolor}

\usepackage{mathrsfs}
\usepackage{multirow}

\allowdisplaybreaks[1]

\def \nn {\nonumber} 

\newcommand{\be}{\begin{equation}}
\newcommand{\ee}{\end{equation}}
\newcommand{\bea}{\begin{eqnarray}}
\newcommand{\eea}{\end{eqnarray}}

 \usepackage{multirow}

\begin{document}

\title{Polar modes {and isospectrality} of Ellis--Bronnikov wormholes}
\author[1]{Bahareh Azad\thanks{\href{mailto:bahareh.azad@uni-oldenburg.de}{bahareh.azad@uni-oldenburg.de}}}
\author[2]{Jose Luis Bl\'azquez-Salcedo\thanks{\href{mailto:jlblaz01@ucm.es}{jlblaz01@ucm.es}}}
\author[3,4]{Xiao Yan Chew\thanks{\href{mailto:xychew998@gmail.com}{xychew998@gmail.com}}}
\author[1]{Jutta Kunz\thanks{\href{mailto:jutta.kunz@uni-oldenburg.de}{jutta.kunz@uni-oldenburg.de}}} 
\author[3,4]{Dong-han Yeom\thanks{\href{mailto:innocent.yeom@gmail.com}{innocent.yeom@gmail.com}}}
\affil[1]{Institute of  Physics, University of Oldenburg, D-26111 Oldenburg, Germany}
\affil[2]{Departamento de F\'isica Te\'orica and IPARCOS, Facultad de Ciencias F\'isicas, Universidad Complutense de Madrid, Spain}
\affil[3]{Department of Physics Education, Pusan National University, Busan 46241, Republic of Korea}
\affil[4]{Research Center for Dielectric and Advanced Matter Physics, Pusan National University, Busan 46241, Republic of Korea}

\date{\today}

\maketitle

\begin{abstract}
We consider polar perturbations of static Ellis--Bronnikov wormholes and derive the coupled set of perturbation equations for the gravitational and the scalar field. 
For massless wormholes the perturbations decouple, and we obtain two identical master equations for the scalar and gravitational modes, which moreover agree with the master equation for the axial modes.
Consequently there is isospectrality with threefold degenerate modes.
{For a finite} mass of the background wormhole solutions, the equations are coupled. We then obtain two distinct branches of polar quasinormal modes for a given multipole number $l$, associated with the presence of the two types of fields.
We calculate the quasi-normal mode frequencies and decay rates for the branches with $l=2, 3$ and 4.  
For a given $l$ the real frequencies of the two branches get the closer, the higher the multipole number gets.
\end{abstract}

\section{Introduction}

The Einstein-Rosen bridge is obtained by solving the vacuum Einstein equations of the general theory of relativity for the static spherically symmetric case \cite{Einstein:1935tc}.
It represents a non-traversable wormhole, encumbered by event horizons.  
Decades later, when considering the Einstein equations in the presence of a scalar field, Ellis \cite{Ellis:1973yv,Ellis:1979bh} and Bronnikov \cite{Bronnikov:1973fh} were able to obtain traversable wormhole solutions, provided they employed a non-standard scalar field, i.e., a phantom field.
The necessity of the presence of a phantom field or, more generally, some form of exotic matter violating the energy conditions in classical general relativity was nicely discussed by Morris and Thorne \cite{Morris:1988cz}, who also contemplated the use of wormholes for rapid interstellar travel, see also \cite{Visser:1995cc,Alcubierre:2017pqm}.

In contrast to classical general relativity, however, alternative theories of gravity do not necessarily require the presence of exotic matter. 
Here the energy conditions can be violated by the gravitational degrees of freedom alone, as it happens, for instance, in Einstein-scalar-Gauss-Bonnet theories \cite{Kanti:2011jz,Kanti:2011yv,Antoniou:2019awm}.
On the other hand, also quantum degrees of freedom like Dirac particles allow for the violation of the energy conditions and thus the emergence of traversable wormholes \cite{Blazquez-Salcedo:2020czn,Blazquez-Salcedo:2021udn,Blazquez-Salcedo:2019uqq,Konoplya:2021hsm}. Recently traversable wormholes have also been constructed in Einstein-3-Form theory {since} the 3-form field can violate the energy conditions {as well \cite{Barros:2018lca,Bouhmadi-Lopez:2021zwt}}.

Clearly, besides the pure theoretical interest %
the various possible ways of detecting a wormhole are of even more interest. 
Considering that wormholes might exist, one of the possibilities to detect them would be optically. 
For instance, one may look for the gravitational lensing effects created by a wormhole 
\cite{Cramer:1994qj,Safonova:2001vz,Perlick:2003vg,Nandi:2006ds,Abe:2010ap,Toki:2011zu,Nakajima:2012pu,Tsukamoto:2012xs,Kuhfittig:2013hva,Bambi:2013nla,Takahashi:2013jqa,Tsukamoto:2016zdu},
one may search for their shadows
\cite{Bambi:2013nla,Nedkova:2013msa,Ohgami:2015nra,Shaikh:2018kfv,Gyulchev:2018fmd,Guerrero:2022qkh,Bouhmadi-Lopez:2021zwt},
seek their accretion disks and radiation associated with quasi-periodic oscillations
\cite{Harko:2008vy,Harko:2009xf,Bambi:2013jda,Zhou:2016koy,Lamy:2018zvj,Deligianni:2021ecz,Deligianni:2021hwt}, {etc.} %

Another interesting possibility for detecting wormholes could arise via studies of scattering some scalar field.
Recently analytic expressions for the transmission and reflection amplitudes of the corresponding effective potential and the absorption cross section of wormholes have been calculated and it has been shown that for both a massless and a massive field, an observer at infinity can easily differentiate between a wormhole and a Schwarzschild black hole by examining the scattering data for the scalar field \cite{Azad:2020ajs}. 

Currently, gravitational wave astronomy \cite{LIGOScientific:2016aoc,LIGOScientific:2017vwq,LIGOScientific:2017ync} is giving us a new potential approach to observe wormholes through their damping modes called quasi-normal modes (see e.g. \cite{Kokkotas:1999bd,Berti:2009kk,Konoplya:2011qq}). 
Quasi-normal modes are characteristic modes of a freely oscillating space-time. 
When a compact object like a black hole, a neutron star or a wormhole oscillates the system is open, and the gravitational waves loses energy and decays in time, so this type of modes is called quasi-normal modes.

Quasi-normal modes of wormholes have been considered before in a variety of contexts \cite{Konoplya:2005et,Kim:2008zzj,Konoplya:2010kv,Konoplya:2016hmd,Volkel:2018hwb,Aneesh:2018hlp,Konoplya:2018ala,Blazquez-Salcedo:2018ipc,Konoplya:2019hml,Churilova:2019qph,Jusufi:2020mmy,Gonzalez:2022ote}.
A particularly interesting aspect here is presented by
the inverse problem, which can allow to find the shape of the wormhole by its quasi-normal modes \cite{Konoplya:2018ala,Volkel:2018hwb}. 
This is different from the case of black holes where a family of effective potentials produces the same quasi-normal mode spectrum \cite{Chandrasekhar:1985kt}.  

A particular example where the quasi-normal modes have allowed to reconstruct the metric near the throat are the symmetric Ellis-Bronnikov wormholes \cite{Konoplya:2018ala}. 
The quasi-normal modes of the general family of Ellis-Bronnikov wormholes have been investigated only partially up to now \cite{Konoplya:2005et,Kim:2008zzj,Blazquez-Salcedo:2018ipc}.
A systematic analysis of the scalar, axial, and radial perturbations was given in \cite{Blazquez-Salcedo:2018ipc}.

Here we calculate the quasi-normal modes for the polar perturbations for the multipole numbers $l=2$, 3 and 4. 
The presence of the scalar field always leads to two such families of modes, as the {mass} %
of the background wormholes is varied, whereas in the %
{massless} case the eigenvalues of the modes coincide.
Indeed, for massless wormholes we obtain two identical master equations for the scalar and metric perturbations.
Moreover, the previously obtained master equation for the axial modes \cite{Blazquez-Salcedo:2018ipc} is also identical to these master equations.
  
The paper is organized as follows. In Section 2 we present the theoretical setting, comprising the action, the equations of motion and the family of static spherically symmetric Ellis-Bronnikov wormhole solutions.
These serve as the background solutions for the perturbations discussed in Section 3, where a brief reminder of the formalism is given, before the equations for the different multipoles, $l=0$, $l=1$, and $l\ge 2$ are obtained.
Additionally it is shown, that the spectrum possesses a threefold degeneracy for the massless wormholes, since in this case all (scalar, polar metric and axial metric)  perturbations satisfy the same master equation.
Starting with a description of the applied methods, we then present our results for the polar quasi-normal modes in Section 4, and we conclude in Section 5.

\section{Theoretical Setting}

We consider the Einstein-Hilbert action
  \be 
  S=\frac{1}{16\pi G}\int d^4x\sqrt{-g}\Big(R+2\nabla_\mu \phi\nabla^{\mu}\phi\Big)
  \ee 
with a massless minimally coupled phantom field $\phi$.
By varying the action we obtain the coupled set of
equations of motion %
  \bea 
   R_{\mu\nu}&=&-2\ \partial_\mu\phi\ \partial_\nu\phi,\\
   \nabla_{\mu}\nabla^{\mu}\phi&=&0.
  \eea 
The family of static spherically symmetric Ellis-Bronnikov wormholes is then given by
  \bea 
  \phi&=& \phi^{(b)}(r) =
  \ \frac{Q}{r_0}\Big[\tan^{-1}\left(\frac{r}{r_0}\right)-\frac{\pi}{2}\Big],\\
   ds^2&=& g_{\mu\nu}^{(b)} dx^{\mu} dx^{\nu} = 
   -e^f dt^2+\frac{1}{e^f}\Big[dr^2+\left(r^2+r_0^2\right)\left(d\theta^2+\sin^2\theta d\varphi^2\right)\Big],
   \eea
with
  \be 
  f=\frac{C}{r_0}\Big[\tan^{-1}\left(\frac{r}{r_0}\right)-\frac{\pi}{2}\Big] \,,
  \ee
  and $Q$, $r_0$, and $C$ are constants discussed below.
  The superscript $(b)$ indicates, that these are the background solutions to be employed in the perturbation expansions.

Asymptotically, for $r \rightarrow +\infty$ the metric function $f$ tends to zero, $f \rightarrow 0$, 
and the metric approaches Minkowski spacetime.  
On the other hand, for $r \rightarrow -\infty$,
a coordinate transformation is needed to approach Minkowski spacetime,
 \begin{equation} \label{coor_trans}
  \bar{t} = e^{-\frac{C \pi}{2 r_0}} t \,, \quad \bar{r} = e^{\frac{C \pi}{2 r_0}} r \,, \quad \bar{r}_0 = e^{\frac{C \pi}{2 r_0}} r_0  \,.
 \end{equation}
Thus the spacetime has two asymptotically flat regions, which are connected by a throat %
where the circumferential radius $R(r)$,
\begin{equation}
R^2(r) = e^{-f}(r^2 + r_0^2) \,,
\end{equation}
assumes its minimal value. %

The global charges of the solutions can be read off from the asymptotic expansion of the phantom field and the metric.
The charge of the phantom field is given by the constant $Q$. 
The mass of the wormhole solution as extracted in the asymptotically flat region $r \to +\infty$ is given by
\begin{equation}
 C =  2 M \,.
\end{equation}

Besides determining the mass, the constant $C$ represents also a measure of the symmetry of the wormhole. 
When $C=0$, the wormhole is massless and symmetric with respect to reflections of $r \to -r$ at the throat $r=0$.
However, when $C \neq 0$, 
the wormhole possesses mass, and it is asymmetric, its throat being located either in the region $r<0$ or $r>0$.
There is also a symmetry relation between solutions with a positive value of $C$ and those with a negative value, which reads
\begin{eqnarray}
f(r,C) = f(-r,-C) - \frac{\pi C}{r_0}, \nonumber \\
\phi(r,C) = -\phi(-r,-C) - \frac{\pi Q}{r_0} \,.
\end{eqnarray}
This relation relates the part of the spectrum of 
quasi-normal modes with $C<0$ with the $C>0$ part.

\section{Perturbation Theory}

   \subsection{Formalism}

We decompose the metric and the phantom field as follows
   \be
   g_{\mu\nu}=g^{(b)}_{\mu\nu}+h_{\mu\nu} \,,
   \ee
    \be
    \phi=\phi^{(b)}+\psi \,.
    \ee
where the superscript ${(b)}$ stands for the background, 
and the perturbations are assumed to be small.
The variation of the Einstein equations with the phantom field source term is
   \be
   \delta R_{\mu\nu}= -2 \delta(\partial_\mu\phi\ \partial_\nu\phi) \,,
   \label{Estn}
   \ee
   where
 \be
  2 \delta R_{\mu\nu}=\nabla_\rho\nabla_\mu\ h^{\rho}_\nu+\nabla_\rho\nabla_\nu\ h^{\rho}_\mu-\nabla_\nu\nabla_\mu\ h-\Box h_{\mu\nu} \,,
 \ee
 and
   \be
   \delta(\partial_\mu\phi\ \partial_\nu\phi)=\partial_\mu\psi\ \partial_\nu\phi^{(b)}+\partial_\mu\phi^{(b)}\ \partial_\nu\psi \,.
   \ee
The variation of the scalar field equation leads to
    \be
    \delta\Box\phi=\Box\psi+\frac{1}{2}\nabla_\lambda h\ \partial^{\lambda}\phi^{(b)}-\nabla_\mu\left(h^{\mu\nu}\partial_\nu\phi^{(b)}\right)=0
    \label{box-phantom}
    \ee 

Employing the Regge-Wheeler gauge \cite{Regge:1957td} the polar perturbations of the metric are given by
    \bea 
    && \hspace*{-1.0cm}
    h^{pol}_{\mu\nu}(t,r,\theta,\phi)= \\
    && \hspace*{-1.0cm}
     \sum\limits_{l}\,\int d\omega
    e^{-i\omega t}P_l(\cos\theta)\left[\begin{array}{cccc}
    e^{f(r)}H_{0l}(r)&H_{1l}(r)&0&0\\
    H_{1l}(r)&e^{-f(r)}H_{2l}(r)&0&0\\
    0&0&e^{-f(r)}(r^2+r_0^2)K_{l}(r)&0\nn\\
    0&0&0&e^{-f(r)}(r^2+r_0^2)\sin^2\theta K_{l}(r)
    \end{array}\right]\,, 
    \eea
    which contains no $m$ dependence because of the spherical symmetry of the background solutions.
Likewise, we employ a spherical harmonic decomposition ($m=0$) of the perturbation $\psi$ of the scalar field, yielding 
    \be 
    \psi(t,r)= \sum\limits_{l}\,\int d\omega
    e^{-i\omega t}u_l(r)\ P_l(\cos\theta)\,.
    \ee 
For given values of $l$ and $\omega$ 
the Laplacian of the phantom field perturbation then becomes
    \be
   \Box\psi=\left(\frac{u_l(r)}{e^{f(r)}} \omega^2
   +e^{f(r)}\left(\frac{2r}{r^2+a^2}\partial_ru_l(r)+\partial^2_r\ u_l(r)-\frac{l(l+1)}{r^2+r_0^2}u_l(r)\right)\right)e^{-i\omega t}P_l(\cos\theta) \,.
   \label{Laplacian_general-psi}
   \ee
   
In the following the index $l$ of the perturbation functions of the metric and the scalar field with be omitted to simplify the notation.
With the scalar spherical harmonics defined for $l\ge 0$, the vector spherical harmonics for $l\ge 1$, and the tensor spherical harmonics for $l\ge 2$, the monopole ($l=0$) case and the dipole ($l=1$) case will be treated in separate subsections, following the quadrupole ($l=2$) case, which will include also the higher multipoles ($l>2$).

\subsection{$l\ge 2$}

We first consider the set of equations to be solved in order to obtain $l\ge 2$ polar quasi-normal modes, since these represent the main objective of the present investigations.
In the derivation of the equations %
we need to distinguish between the two cases $C\neq 0$ and $C=0$ which are treated consecutively in the following two subsubsections.

\subsubsection{$C\neq 0$}

From the Einstein field equations we obtain seven nontrivial equations, associated with the perturbations of the Ricci tensor $\delta R_{tt}$, $\delta R_{tr}$, $\delta R_{t\theta}$ $(\delta R_{t\varphi})$, $\delta R_{rr}$, $\delta R_{r\theta}$ $(\delta R_{r\varphi})$, $\delta R_{\theta\theta}$ and $\delta R_{\varphi\varphi}$. 
We extract $K''$ from $\delta R_{\theta\theta}$ and replace it in $\delta R_{\varphi\varphi}$. 
We then find the equality
   \be 
   H_2=H_0\,.
   \ee 
Next we eliminate $H_2$ in the remaining six equations and obtain
   \bea 
   \delta R_{tt}&=&\Bigg(-\frac{e^{2f}}{2}H_0''-\frac{e^{2f}\left(C+2r\right)}{2\left(r^2+r_0^2\right)}H_0'+\left(\frac{l(l+1)e^{2f}}{2(r^2+r_0^2)}+\frac{\omega^2}{2}\right)H_0,\nn\\
   &+&i\omega e^{2f}\left(-H_1'+\frac{C-4r}{2(r^2+r_0^2)}H_1\right)+\frac{Ce^{2f}}{2(r^2+r_0^2)}K'+\omega^2 K\Bigg)\ e^{-i\omega t}P_l(\cos\theta) 
   =0\,,
   \label{DeltaR_ttC}
   \eea 
    \bea 
    \delta R_{tr}&=&\Bigg(-i\omega\left(\frac{C-2r}{2(r^2+r_0^2)}H_0+\frac{C-r}{r^2+r_0^2}K-K'\right)+\frac{l(l+1)e^{2f}}{2(r^2+r_0^2)}H_1\Bigg)\ e^{-i\omega t}P_l(\cos\theta) \nn \\
    &=&2i\omega\ e^{-i\omega t}\ u(r)\ P_l(\cos\theta)\ \partial_r\phi^{(b)}
    \,,
    \label{DeltaR_trC}
    \eea 
   \be
   \delta R_{t\theta}=\frac{1}{2}\Bigg(i\omega(H_0+K)+e^{2f}\left(\frac{C}{r^2+r_0^2}H_1+H_1'\right)\Bigg)\ e^{-i\omega t}\ \partial_{\theta}\ P_l(\cos\theta) =0\,,
   \label{DeltaR_tthetaC}
   \ee 
    \bea 
    \delta R_{rr}&=&\Bigg(\frac{1}{2}\left(\left(\frac{l(l+1)}{r^2+r_0^2}-\omega^2e^{2f}\right)H_0+\frac{C+2r}{r^2+r_0^2}H_0'+H_0''\right)\nn\\
    &+&i\omega e^f\left(\frac{C}{2(r^2+r_0^2)}H_1+H_1'\right)+\frac{C-4r}{2(r^2+r_0^2)}K'-K''\Bigg)\ e^{-i\omega t}P_l(\cos\theta) \nn \\
    &=&-4\ e^{-i\omega t}\ \partial_ru(r)\ P_l(\cos\theta)\ \partial_r\phi^{(b)}
    \,,
    \label{DeltaR_rrC}
    \eea
    \bea
    \delta R_{r\theta}&=&\frac{1}{2}\Bigg(\frac{C}{r^2+r_0^2}H_0+H_0'+\omega e^{-f}H_1-K'\Bigg)\ e^{-i\omega t}\partial_{\theta}P_l(\cos\theta) \nn \\
    &=&-2\ e^{-i\omega t}\ u(r)\ \partial_{\theta}P_l(\cos\theta)\ \partial_r\phi^{(b)}
    \,,
    \label{DeltaR_rthetaC}
    \eea
and 
   \bea
   \delta R_{\theta\theta}&=& \Bigg(H_0-\frac{C-2r}{2}H_0'+\frac{i\omega}{2}(C-2r)e^{-f}H_1+\frac{1}{2}\left((l-1)(l+2)-\omega^2(r^2+r_0^2)e^{-2f}\right)K\nn\\
   &+&\frac{1}{2}(C-4r)K'-\frac{1}{2}(r^2+r_0^2)K''\Bigg)\ e^{-iwt}P_l(\cos\theta) 
   =0\,.
   \label{DeltaR_thetathetaC}
   \eea  
The equations \eqref{DeltaR_trC}, \eqref{DeltaR_tthetaC} and \eqref{DeltaR_rthetaC} are of first order.
We now solve the equation of $\delta R_{tr}$ for $K'(r)$,  the equation of $\delta R_{t\theta}$ for $H'_1(r)$, and the equation of $\delta R_{r\theta}$ for $H'_0(r)$ after inserting $K'(r)$ here.
This leads to
   \be 
   K'=\frac{1}{r^2+r_0^2}\Bigg(-\frac{C-2r}{2}H_0+\frac{l(l+1)}{2\omega}e^fH_1+(C-r)K+\sqrt{C^2+4r_0^2}\ u(r)\Bigg) \,,
   \label{K1p}
   \ee 
   \be 
   H_1'=-i\omega e^{-f}\left(H_0+K\right)-\frac{C}{r^2+r_0^2}H_1 \,,
   \label{eqH1}
   \ee 
and 
   \be
   H_0'=\frac{1}{r^2+r_0^2}\Bigg(\frac{2r-3C}{2}H_0+\frac{i}{2}\left(\frac{l(l+1)}{\omega}e^f-2\omega(r^2+r_0^2)e^{-f}\right)H_1+(C-r)K-\sqrt{C^2+4r_0^2}\ u(r)\Bigg)  \,.
   \label{eqH0}
   \ee 
Now we extract $K''(r)$ from $\delta R_{\theta\theta}$ and $H''_0(r)$ from $\delta R_{tt}$ and insert these together with the first derivatives of the metric functions in $\delta R_{rr}$. This leads to the \textit{algebraic relation}
   \bea
   & &\sqrt{C^2+4r_0^2}\ \Big(\frac{2u'}{r^2+r_0^2}-\frac{u\left(C-4r\right)}{\left(r^2+r_0^2\right)^2}\Big)
   +\Bigg(\frac{2\left(l-1\right)\left(l+2\right)}{r^2+r_0^2}
   -\frac{3C\left(C-2r\right)}{\left(r^2+r_0^2\right)^2}\Bigg)\ H_0 \nn\\
   &+& i\ \Bigg(\frac{2\omega\left(C-2r\right)}{r^2+r_0^2}\ e^{-f}+\frac{l\left(l+1\right)C}{\omega\left(r^2+r_0^2\right)^2}\ e^{f}\Bigg) \ H_1 \nn\\
   &+&2\Bigg(2\omega^2\ e^{-2f}-\frac{(l-1)(l+2)}{r^2+r_0^2}
   +\frac{C(C-r)}{(r^2+r_0^2)^2}\Bigg)\ K=0.
   \label{Al}
   \eea

When solving this \textit{algebraic relation} for $u'$, we are left with four first order equations for the four variables.
By taking the first and second derivative of the \textit{algebraic relation} we obtain the three metric perturbation functions $H_0$, $H_1$, and $K$ in terms of $u$ and its derivatives.
If we would set $u(r)=0$ (i.e., turn off the scalar perturbation) the metric perturbations would also vanish. 
This means that the metric perturbations are coupled with the phantom field perturbation, and we can not decouple the phantom field from metric perturbations, in general.
We next need to study the behavior of the perturbation functions at both radial infinities.
We therefore make series expansions for the functions assuming asymptotic flatness.
Employing the tortoise coordinate $r^*$ %
   \be
   \frac{dr^*}{dr}=e^{-f}
   \label{tortoise}
   \ee
the expansion for $r \to +\infty$ becomes
     \bea  
     u(r)&=&e^{i\omega r^*}\Big(\frac{u_1}{r}+\frac{u_2}{r^2}+\frac{u_3}{r^3}+...\Big) \,,
     \label{onlyu}\\
     H_0(r)&=&e^{i\omega r^*}\Big(i\omega K_0 r+\frac{\left(3iC\omega -(l{+}2)(l{-}1)\right)}{2}K_0+...\Big) \,,\\
     H_1(r)&=&e^{i\omega r^*}\Big(-i\omega  K_0 r -\frac{\left(3iC\omega-(l{+}2)(l{-}1)\right)}{2}K_0+...\Big) \,,\\
     K(r)&=&e^{i\omega r^*}\Big(K_0-\frac{\omega^2(C^2+4r_0^2)+6iC\omega -l^4-2l^3+l^2+2l}{8\omega^2 r^2}K_0+...\Big) \,,
     \label{onlyK}
     \eea 
where {$u_1=A_s^+$ and $K_0=A_g^+$} {are free amplitudes that fix the rest of the parameters of the expansion}
     \bea  
     u_2&=&\frac{1}{4\omega}  \Big(\left(\sqrt{C^2+4r_0^2}K_0 -2u_1 \right)C\omega+2i l(l+1)u_1\Big) \,,\\
     u_3&=& \frac{1}{8\omega^2}\Big( 
     \left(
     \omega^2(C^2-4r_0^2)  - 2i\omega C (2l^2+2l-1) -l^4-2l^3+l^2+2l
     \right) u_1
     \nn
     \\
     &+& i\omega C \sqrt{C^2+4r_0^2} (l+2)(l-1) K_0 \Big) \, .
     \eea       
On the other hand, for $r \to -\infty$ the expansion becomes
     \bea  
     \bar{u}(r)&=&e^{-i\omega r^*}\Big(\frac{\bar{u}_1}{r}+\frac{\bar{u}_2}{r^2}+\frac{\bar{u}_3}{r^3}+...\Big) \,,
     \label{baru}\\
     \bar{H}_0(r)&=&e^{-i\omega r^*}\Big(-i\omega r e^{C\pi/r_0}\bar{K}_0 -\frac{3iC
     \omega  e^{C\pi/r_0}+(l+2)(l-1)}{2}\bar{K}_0+...\Big) \,,\\
     \bar{H}_1(r)&=&e^{-i\omega r^*}\Big(-i\omega r e^{C\pi/r_0}\bar{K}_0 -\frac{3iC
     \omega  e^{C\pi/r_0}+(l+2)(l-1)}{2}\bar{K}_0+...\Big) \,,\\
     \bar{K}(r)&=&e^{-i\omega r^*}\Big(\bar{K}_0-\frac{\omega^2(C^2+4r_0^2)-6iC\omega e^{-C\pi/r_0} -(l^4+2l^3-l^2-2l)e^{-2C\pi/r_0}}{8\omega^2 r^2}\bar{K}_0+..\Big) \,,
     \label{barK}
     \eea 
where again {$\bar u_1=A_s^-$ and $\bar K_0=A_g^-$} {are free amplitudes and}
     \bea  
     \bar{u}_2&=&\frac{1}{4\omega}  \Big(\left(\sqrt{C^2+4r_0^2}\ \bar{K}_0-2\bar{u}_1\right) C \omega-2il(l+1)e^{-C\pi/r_0}\bar{u}_1\Big) \,,\\
     \bar{u}_3&=& \frac{1}{8\omega^2}\Big( 
     \left(
     \omega^2(C^2-4r_0^2)  + 2i\omega e^{-C\pi/r_0} C (2l^2+2l-1) - (l^4+2l^3-l^2-2l)e^{-2C\pi/r_0}
     \right) \bar{u}_1
     \nn
     \\
     &+& i\omega e^{-C\pi/r_0} C \sqrt{C^2+4r_0^2} (l+2)(l-1) \bar{K}_0 \Big) \, .     
     \eea

\subsubsection{$C=0$}

In the massless case the first order equations reduce to
    \bea 
    K'&=& \frac{1}{2\omega\left(r^2+r_0^2\right)}\Big(il\left(l+1\right)H_1-2\omega\left(rK-rH_0-2r_0u\right)\Big) \,, \label{KC0} \\
     H_0'&=& \frac{1}{2\omega\left(r^2+r_0^2\right)} \Big(\left(-2i\omega^2\left(r^2+r_0^2\right)+il\left(l+1\right)\right)H_1-2\omega\left(rK-rH_0+2r_0u\right)\Big) \,, \label{H0C0} \\
    H_1'&=&-i\omega\left(H_0+K\right) \,, \label{H1C0} \\
    u'&=&-\frac{1}{4r_0}\Big(\left(2\omega^2\left(r^2+r_0^2\right)+(l+2)(l-1)\right)K+\left(l+2\right)\left(l-1\right)H_0-2i\omega rH_1\Big)-\frac{2r}{r_0^2+r^2}u \,. \,\,\,\,\,\,\,\,\,
    \label{upC0} \eea
The asymptotic expansions shown in the previous section remain valid when we substitute $C=0$.

Calculating $u''$ by taking the derivative of (\ref{upC0}) leads to a decoupled equation for the scalar perturbation,
    \be 
    u''=-\frac{2r}{r_0^2+r^2}u'-\frac{1}{(r_0^2+r^2)^2}\Big(r_0^4\omega^2+\left(2r_0^2\omega^2-l^2-l+4\right)r_0^2+\left(r^2\omega^2-l^2-l\right)r^2\Big)u \,.
    \label{eq_scalar_C0}
    \ee 
A non-trivial solution to this equation 
can be written in closed form:
    \bea 
    u(r)
    &=&\frac{C_1\ r}{r_0^2+r^2} \ \textrm{HeunC}\Big(0,\frac{1}{2},-2,-\frac{r^2_{{0}}\omega^2}{4},\frac{1}{4}\left(r_0^2\omega^2-l^2-l+5\right),-\frac{r^2}{r_0^2}\Big)\nn\\
    &+&\frac{C_2}{r_0^2+r^2} \ \textrm{HeunC}\Big(0,-\frac{1}{2},-2,-\frac{r^2_{{0}}\omega^2}{4},\frac{1}{4}\left(r_0^2\omega^2-l^2-l+5\right),-\frac{r^2}{r_0^2} 
    \Big) \,,
    \eea
where $\textrm{HeunC}$ is the Heun Confluent function, $C_1$ and $C_2$ are constants. 

The equations for the space-time perturbations can be cast into a single second order differential equation for $H_1$,
     \bea   
     &&H_1'' = \frac{8i\omega r_0 r u' + 16i\omega r_0 r^2 u /(r^2+r_0^2)-2r\omega^2\left(r^2+r_0^2\right)H_1'}{\left(l^2+l-2-\left(r^2+r_0^2\right)\omega^2\right)\left(r^2+r_0^2\right)} \nn \\
     &&+ \frac{\Big(\left(r^2+r_0^2\right)^2\omega^4+\left[6r^2+2r_0^2-2\left(r^2+r_0^2\right)l\left(l+1\right)\right]\omega^2+l^4+2l^3-l^2-2l\Big)H_1}{\left(l^2+l-2-\left(r^2+r_0^2\right)\omega^2\right)\left(r^2+r_0^2\right)} 
    \label{eq_d2H1_C0}
     \eea

The previous perturbation equations can be written in terms of a single master equation, that coincides with the master equation of the axial perturbations. First note that, if we parametrize the scalar perturbation in terms of the master variable $Z(r)$ such as
   \be   
   u=\frac{1}{\sqrt{r^2+r_0^2}} Z(r),
   \ee   
then equation (\ref{eq_scalar_C0}) can be written as a
Schr\"odinger-like master equation  
   \be   
   \frac{d^2Z(r)}{dr^2}+\left(\omega^2-V(r)\right)Z(r)=0,
\label{eq_Schr}
   \ee 
with effective potential, 
    \be  
    V(r)=\frac{l\left(l+1\right)}{r^2+r_0^2}-\frac{3r_0^2}{(r^2+r_0^2)^2}
    \ee 
Note that, for $C=0$ the tortoise coordinate (\ref{tortoise}) $r^*$ is the same as $r$. 
This potential is the same as the one found in \cite{Blazquez-Salcedo:2018ipc} for the axial perturbations with $C=0$.

It is also possible to cast the space-time perturbation equation (\ref{eq_d2H1_C0}) into the same master equation. However the transformation is more involved. We define
     \bea   
    H_1(r)&=&A(r)\hat Z(r),\\
    u(r)&=&B(r)\hat Z(r),
    \eea   
where $\hat Z(r)$ is the new master variable and $A(r)$ and $B(r)$ are 
\bea
&& A(r) = \frac{(r^2+r_0^2)^{\sqrt{3}-3/2}}{l^2+l-2-\omega^2(r^2+r_0^2)}  \Big[  \\
&& D_1 r^3 
\textrm{HeunG}\Big(
-r_0^2\omega^2\beta,
-\frac{r_0^2\omega^2\beta}{2}+\frac{5\sqrt{3}}{2}+3,
\frac{5}{4}+\frac{\sqrt{149}}{4}+\sqrt{3},
\sqrt{3}+\frac{5}{4}-\frac{\sqrt{149}}{4},
\frac{5}{2},
0,
\beta\omega^2r^2
\Big) \nn \\
&& +D_2 
\textrm{HeunG}\Big(
-r_0^2\omega^2\beta,
-\frac{r_0^2\omega^2\beta}{2}-\frac{\sqrt{3}}{2}+\frac{3}{2},
-\frac{1}{4}+\frac{\sqrt{149}}{4}+\sqrt{3},
\sqrt{3}-\frac{1}{4}-\frac{\sqrt{149}}{4},
-\frac{1}{2},
0,
\beta\omega^2r^2
\Big)
\Big] \, , \nn \\
&& B(r) = -\frac{i\omega(r_0^2+r^2)}{4r_0}A - \frac{i(r_0^2+r^2)(l^2+l-2-\omega^2(r_0^2+r^2))}{4r_0r\omega}A' \, ,
\eea
where $\beta^{-1}=l^2+l-1-r_0^2\omega^2$ and $\textrm{HeunG}$ is the Heun General function, $D_1$ and $D_2$ are two arbitrary constants.

Then it is possible to show that equation (\ref{eq_d2H1_C0}) simplifies into equation (\ref{eq_Schr}) for $\hat Z(r)$ and with the same axial perturbations potential. As we will explicitly calculate later, this means that the spectrum of polar quasinormal modes of the $C=0$ wormholes is exactly the same as the axial spectrum.

\subsection{$l=1$} 

For $l=1$ three of the Einstein's equations are identically zero,
    \be 
    \delta R_{t\varphi}=\delta R_{r\varphi}=\delta R_{\theta\varphi}=0 \,.
    \ee 
From the non-trivial equations we obtain the following three first order differential equations,
    \bea 
         \label{K}
     K' &=& \frac{1}{r_0^2+r^2}\Big(\frac{ie^{f}}{\omega}H_1-\frac{C-2r}{2}H_2+\left(C-r\right)K+\sqrt{C^2+4r_0^2}\ u\Big)\,,\\
      \label{F1}
     H_1' &=& -\frac{C}{r^2+r_0^2}H_1-i\omega e^{-f} \left(H_2+K\right) \,,\\    
     H_0' &=&-\frac{C-r}{r^2+r_0^2}H_0-i\Big(\omega e^{-f}-\frac{e^{f}}{\omega\left(r^2+r_0^2\right)}\Big)H_1\nn\\
     &-&\frac{C}{2\left(r^2+r_0^2\right)}H_2+\frac{C-r}{r^2+r_0^2}K-\frac{\sqrt{C^2+4r_0^2}}{r^2+r_0^2} u \,,     \label{F0}
     \eea
and the following \textit{algebraic equation},
    \bea
    &-&\frac{C^2-3Cr-2r_0^2}{2(r^2+r_0^2)}H_0+\frac{i}{2}\left(\frac{C}{\omega(r^2+r_0^2)^2}+\frac{w(C-2r)e^{-f}}{(r^2+r_0^2)}\right)H_1-\frac{c^2+4r_0^2}{4(r^2+r_0^2}H_2\nn\\
    &+&\left(\omega^2e^{-f}+\frac{C(C-r)}{2(r^2+r_0^2)^2}\right)K+\frac{\sqrt{C^2+4r_0^2}}{r^2+r_0^2}\left(\frac{4r-C}{2(r^2+r_0^2)}u+u'\right)=0. \label{alg_l1}
    \eea
This set of equations still has a gauge freedom that we can use to simplify the system further. 
Let us discuss here two possibilities. 

Imposing the gauge $H_1=0$, we obtain from $\delta R_{t\theta}$ 
     \be 
     K=-H_2 \,.
     \ee 
Then it is possible to simplify the remaining perturbation equations and obtain the following system of first order differential equations
    \bea 
     H_2'&=&\frac{(4r-3C)H_2+2\sqrt{C^2+4r_0^2}\ u
}{2(r^2+r_0^2)}\,,  \\
     H_0'&=&\frac{1}{r^2+r_0^2}\Big(\left(r-C\right)H_0+\frac{2r-3C}{2}H_2-\sqrt{C^2+4r_0^2}\ u\Big) \,,
     \eea
and 
    \bea 
     u'&=&\frac{C^2-3Cr-2r_0^2}{2\left(r^2+r_0^2\right)\sqrt{C^2+4r_0^2}}H_0\nn\\
     &+&\frac{1}{\sqrt{C^2+4r_0^2}}\Big(\omega^2 e^{-f}+\frac{r_0^2}{r^2+r_0^2}+\frac{3C^2-2Cr}{4\left(r^2+r_0^2\right)}\Big)H_2\nn\\
     &+&\frac{C-4r}{2\left(r^2+r_0^2\right)}\ u \,.
    \eea 
We are then left with three first order equations to calculate the quasi-normal modes.

Alternatively, we may consider the variation of the phantom field (\ref{box-phantom})), which leads to the second order equation for the function $u$
   \be
   u''=\frac{\sqrt{C^2+4r_0^2}}{4(r^2+r_0^2)}\left(H_0'+2i\omega e^{-f}H_1+H_2'-2K'\right)-\left(\omega^2e^{-2f}+\frac{2}{r^2+r_0^2}\right)u-\frac{2{r}}{r^2+r_0^2}u' \,.
   \ee
This equation is compatible with (\ref{K}-\ref{alg_l1}).
We may now choose the gauge
   \be 
   H_0'+2i\omega e^{-f}H_1+H_2'-2K'=0 \,,
   \ee 
to simplify the equation, yielding
   \be
   u''=-\left(\omega^2e^{-2f}+\frac{2}{r^2+r_0^2}\right)u-\frac{2{r}}{r^2+r_0^2}u'\,.
   \label{eq_sc_l1}
   \ee
This is the equation for $l=1$ scalar field perturbations in the background of the Ellis wormhole as shown in %
\cite{Blazquez-Salcedo:2018ipc}, from which we can calculate the scalar quasinormal modes.

\subsection{$l=0$}

In this case %
the function $H_1$ does not contribute in the Einstein equations. 
Thus for simplicity we put $H_1=0$. 
We now again consider the variation of the phantom field (\ref{box-phantom})), yielding for $u$ the equation
     \be  
     u''=\frac{\sqrt{C^2+4r_0^2}}{4(r^2+r_0^2)}\left(H_0'+H_2'-2K'\right)-\omega^2 e^{-2f}\ u -\frac{2r}{r^2+r_0^2}u' \,.
     \ee         
We now fix the gauge as follows
   \bea  
   H_0&=&2K-H_2 \,.
   \eea 
The scalar field equation decouples from space-time perturbations
     \be  
     u''=-\omega^2 e^{-2f}\ u -\frac{2r}{r^2+r_0^2}u' \,.
        \label{eq_sc_l0}
     \ee  
Non-trivial solutions to this equation result in the spectrum of $l=0$ scalar quasinormal modes.
The space-time perturbations can be cast into a second order differential equation for $K$,
     \bea  
     K''&=&-\frac{(2Cr+4r_0^2)K'}{(r^2+r_0^2)(C-2r)} + \frac{2CK}{(r^2+r_0^2)(C-2r)} - \omega^2 e^{-2f} K + \frac{4u\sqrt{C^2+4r_0^2}}{(r^2+r_0^2)(C-2r)} \,.
     \eea 
The unstable modes are found for solutions with $u=0$. 
With help of the function $Z$,
      \be       
      K=\frac{(C-2r)}{\sqrt{r^2+r_0^2}}\ e^{\frac{C}{2r_0}\arctan(\frac{r}{r_0})}\ \ Z \,,\\
      \ee 
and the tortoise coordinate $r^*$ (equation (\ref{tortoise}))
we rewrite the equation into a Schr\"odinger-like equation,
     \be   
    \frac{d^2Z}{dr^*}+\left(\omega^2-V_r(r)\right)Z=0 \,.
    \label{eq_rad_pert}
    \ee 
with the effective potential $V_r(r)$
    \be    
   V_r= \frac{e^{2f}}{4(C-2r)^2(r^2+r_0^2)}\Big(C^4-8rC^3+12(r^2-r_0^2)C^2-16r(r^2-r_0^2)C-16r_0^2(3r^2+2r_0^2)\Big) \,.
    \ee
     
The above expression corrects the misprints in the corresponding effective potential in \cite{Blazquez-Salcedo:2018ipc}.
Since the effective potential has a singularity at $C=2r$, we study its behaviour near this singularity
    \be     
    V_r(r \to \frac{C}{2})=
    \frac{2 e^{\frac{2C}{r_0}\left(\tan^{-1}(\frac{C}{2r_0})-\frac{\pi}{2}\right)}}{\left(r-\frac{C}{2}\right)^2}
    +\frac{8Ce^{\frac{2C}{r_0}\left(\tan^{-1}(\frac{C}{2r_0})-\frac{\pi}{2}\right)}}{\left(C+4 r_0^2\right)\left(r-\frac{C}{2}\right)}
    -\frac{4e^{\frac{2C}{r_0}\left(\tan^{-1}(\frac{C}{2r_0})-\frac{\pi}{2}\right)}}{C^2+4r_0^2}+... \,.
    \ee     
For $r \to +\infty$ the effective potential becomes
    \be    
    V_r(r \to +\infty)= \frac{C}{r^3}-\frac{-\frac{7C^2}{4}+3r_0^2}{r^4}+... \,,
    \ee    
and for $r \to -\infty$ 
     \be    
      V_r(r \to -\infty)=\frac{ e^{-\frac{2\pi C}{r_0}}C}{r^3}-\frac{e^{-\frac{2\pi C}{r_0}}C^2+\frac{e^{-\frac{2\pi C}{r_0}}\left(12C^2-48r_0^2\right)}{16}}{r^4}+... \,.
     \ee  
     \begin{figure}
  \centering
  \includegraphics[angle =-90,width=0.48\textwidth]{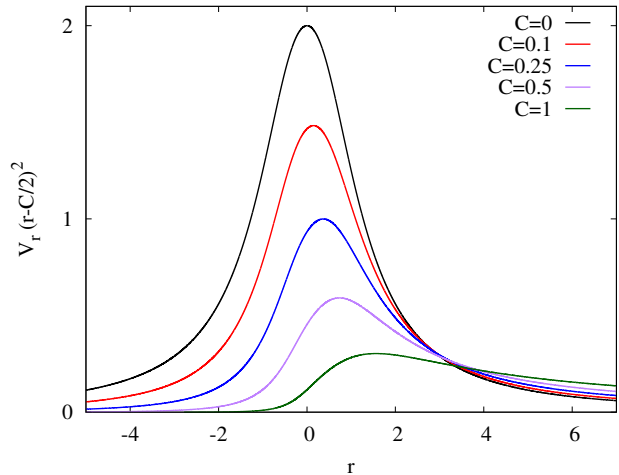}
  \caption{
  Effective potential $V_r(r) \cdot (r-C/2)^2$ versus radial coordinate $r$ for 
  $C=0$, $0.1$, $0.25$, $0.5$ and $1$ ($r_0=1$).}
  \label{Fig:potential}
  \end{figure}
 We exhibit the effective potential $V_r(r)$, multiplied by the factor $(r-C/2)^2$ to cancel the singularity, in figure \ref{Fig:potential} for $C = 0$, 0.1, 0.25, 0.5 and 1 ($r_0=1$).
 We note, that this figure corrects the corresponding figure in \cite{Blazquez-Salcedo:2018ipc}. The calculations of the perturbation equations and the quasi-normal modes were, however, not affected by the misprints.

\section{Results}

We now study the spectrum of polar quasi-normal modes.
Since these characteristic modes of oscillation are damped, the modes are complex,
   \be  
   \omega=\omega_R+i\omega_I,
   \ee   
with frequency $\omega_R$ and decay rate $\omega_I$ (and sign convention $\omega_I<0$ for stable modes). 
In the following we first discuss the method used to calculate the quasi-normal modes and present the numerical results for $l\ge 2$.
Subsequently we briefly recall the method and results for $l=1$ and $l=0$.

\subsection{$l\ge 2$}

\subsubsection{Method}

For the numerical calculation of the quasinormal modes with $l\ge 2$, we rewrite the system of equations \ref{K1p}) - (\ref{Al}). %
This system can be simplified into a second order ordinary differential equation (ODE) for $u$, coupled with two first order ODEs  for the functions $H_1$ and $K$ (the other perturbation functions are then given algebraically in terms of these three functions).
Schematically this set may be expressed as 
\be
\frac{d}{dr}{\vec{Z}}+\mathbf{M}\vec{Z}=0\,,
\label{eq:perturbations}
\ee
where $\vec Z$ denotes the column vector with components $u$, $u'$, $H_1$ and $K$. 
The matrix $\mathbf{M}$ contains the coupling among the perturbation functions and the background functions. 

To obtain the quasi-normal modes we need to solve the coupled set of equations
\eqref{eq:perturbations} subject to the corresponding set of physically motivated boundary conditions,
i.e., we have to impose that we do not have any incoming waves from infinity. Therefore the modes are purely outgoing at infinity, $r^* \to \pm\infty$ (see (\ref{onlyu})-(\ref{onlyK}) and (\ref{baru})-(\ref{barK})), with components $Z_i$
\be Z_i \sim \left\{ 
\begin{array}{cc}
 e^{i \omega r^*}\,, &
 \ \ \ \ r\to + \infty \,, \\
 e^{-i \omega r^*}\,, &
 \ \ \ \ r\to - \infty \,.
 \end{array} \right.
\ee

For the numerical integration we divide space at some value $r_c$ into two regions.
In the region $r>r_c$, perturbation function $Z_i(r)$ possess the asymptotic behaviour for $r \to +\infty$ \cite{Chandrasekhar:1985kt}
\begin{align}
r>r_c \,, \quad  Z_i^+(r) &=  e^{i \omega r^{*}} Z_i^P(r)  \,,
\end{align}
while in the region $r<r_c$ their asymptotic behaviour for $r \to -\infty$ is given by
\begin{align}
r<r_c \,, \quad  Z_i^-(r) &=  e^{-i \omega r^{*}} Z_i^N(r)  \,.
\end{align}
Based on the corresponding expansions for $r \to \pm \infty$, we then generate independent solutions (with different values for the scalar and gravitational amplitudes $A_s^\pm$ and $A_g^\pm$) for the functions $Z_i^N(r)$ and $Z_i^P(r)$ for some chosen value of $\omega$.
Then we match these functions at $r=r_c$ and calculate the derivatives of the functions.
We obtain the quasi-normal mode, when a linear combination of the two independent solutions in one region smoothly matches a linear combination of the solutions of the other region (see subsection 4.1.3 for further details).

In order to integrate numerically the equations subject to the corresponding boundary conditions, we use the package Colsys \cite{Ascher:1979iha}, a collocation method for systems of ordinary differential equations with error estimation and adaptive mesh selection.

\subsubsection{Spectrum}

For $l\ge 2$ we have  two families of modes \cite{Blazquez-Salcedo:2016enn}.
For black holes they can be labeled as
gravitational-led modes, that are dominated by the gravitational perturbations, i.e., their dominant amplitude is $A_g^{\pm}$, and scalar-led modes with dominant amplitude $A_s^{\pm}$.
Here such a clear distinction seems not possible.
Therefore we refer to the two families of modes for a given $l$ as branch 1 and branch 2, {where we currently focus on the fundamental branches}.
{However, we might obtain such a classification if we were to consider larger wormhole masses, since in this limit the wormhole modes are expected to approach those of the Schwarzschild black hole \cite{Blazquez-Salcedo:2018ipc}.
}

\begin{figure}
  \centering
  \includegraphics[angle =-90,width=0.48\textwidth]{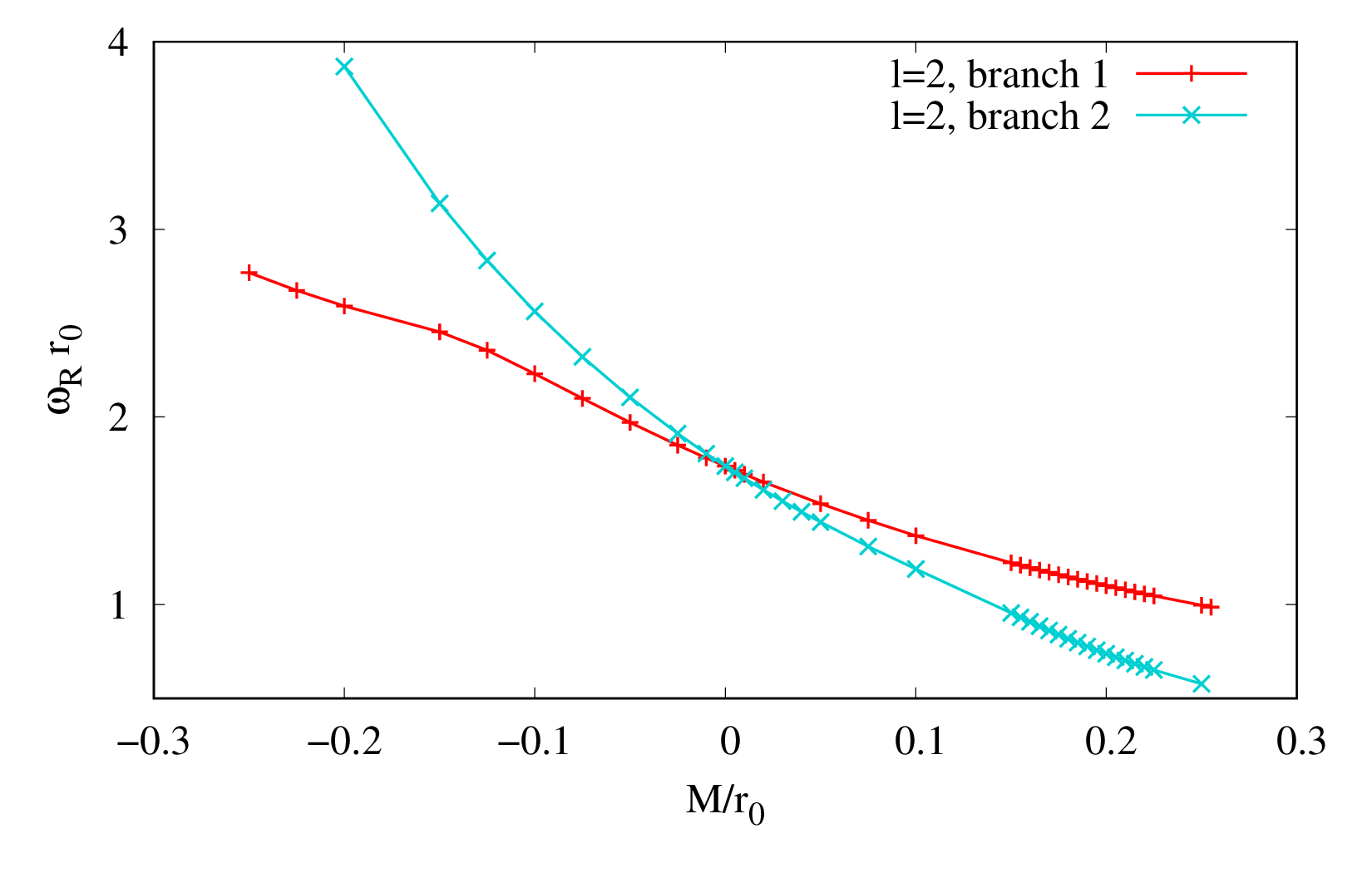}
   \includegraphics[angle =-90,width=0.48\textwidth]{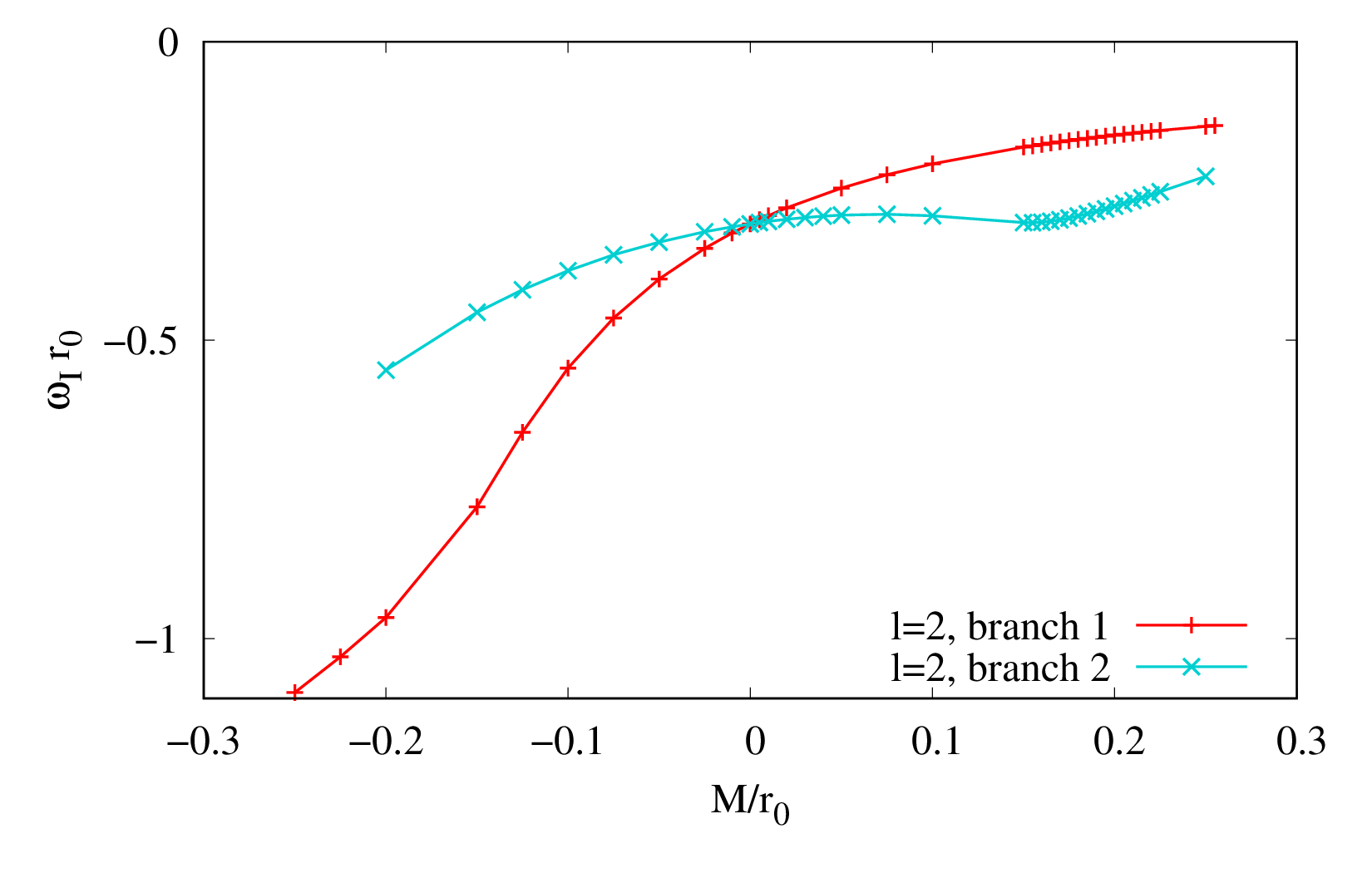}
  \caption{
  Polar $l=2$ quasi-normal modes: dimensionless frequency $\omega_R r_0$ (left) and dimensionless decay rate $\omega_I r_0$ (right) vs dimensionless mass $M/r_0$ ($r_0=1$). 
  \label{Fig:polar_R_l2}}
  \end{figure}

We exhibit the two fundamental branches of polar quadrupole ($l=2$) modes in figure \ref{Fig:polar_R_l2}. Here the left figure shows the scaled frequency $\omega_R r_0$ versus the scaled mass $M/r_0$, while the right figure shows the scaled decay time $\omega_I r_0$ versus the scaled mass.
We note, that the two branches cross precisely at $M=0$, {since} the eigenvalue $\omega$ degenerates in the massless case.
Away from the crossing, the frequencies and decay times are quite distinct for the two branches. A selection of values for these quasi-normal modes is also reported in Table \ref{table1}.

\begin{table}
\centering
\begin{tabular}{|c|c|c|c|c|c|c|c|} 
\hline
\textbf{$M/r_0$}       &   Branch     & \multicolumn{2}{c|}{$l$=2}               & \multicolumn{2}{c|}{$l$=3}                       & \multicolumn{2}{c|}{$l$=4}                        \\ 
\hline
                      &        & \textbf{$\omega_R$}               & \textbf{$\omega_I$}                & \textbf{$\omega_R$}               & \textbf{$\omega_I$}                & \textbf{$\omega_R$}               & \textbf{$\omega_I$}                 \\ 
\hline
\multirow{2}{*}{0.2}  & 1 & 1.10                  & -0.157                  & 1.76                  & -0.199                  & 2.39                  & -0.223                   \\
                      & 2 & 0.738                  & -0.275                  & 1.70                  & -0.330                  & 2.35                  & -0.320                   \\
\hline
\multirow{2}{*}{0.15}  & 1 & 1.22                  & -0.177                  & 1.98                  & -0.230                  & 2.69                  & -0.258                   \\
                      & 2 & 0.955                 & -0.303                  & 1.91                  & -0.348                  & 2.66                  & -0.346                   \\
\hline
\multirow{2}{*}{0.1}  & 1 & 1.37                  & -0.205                  & 2.24                  & -0.271                  & 3.06                  & -0.305                   \\
                      & 2 & 1.19                  & -0.292                  & 2.18                  & -0.364                  & 3.03                  & -0.375                   \\
\hline                      
\multirow{2}{*}{0.05}  & 1 & 1.54                  & -0.245                  & 2.56                  & -0.329                  & 3.51                  & -0.367                   \\
                      &  2 & 1.44                  & -0.290                  & 2.52                  & -0.382                  & 3.50                  & -0.408                   \\
\hline                      
{0}    
& 1 $\&$ 2 & 
{1.74} & {-0.305} & {2.95} & {-0.410} & {4.08} & {-0.449}  \\
\hline                      
\multirow{2}{*}{-0.05} & 1 & 1.97                  & -0.398                  & 3.45                  & -0.524                  & 4.78                  & -0.559                   \\
                      & 2 & 2.10                  & -0.336                  & 3.50                  & -0.451                  & 4.81                  & -0.502                   \\
\hline                      
\multirow{2}{*}{-0.1} & 1 & 2.23                  & -0.547                  & 4.08                  & -0.681                  & 5.68                  & -0.703                   \\
                      &  2 & 2.56                  & -0.384                  & 4.20                  & -0.509                  & 5.73                  & -0.571                   \\
\hline                      
\multirow{2}{*}{-0.15} & 1 & 2.45                  & -0.779                  & 4.90                  & -0.894                  & 6.82                  & -0.888                   \\
                      & 2 & 3.14                  & -0.453                  & 5.07                  & -0.588                  & 6.90                  & -0.663                   \\
\hline                      
\multirow{2}{*}{-0.2} & 1 & 2.59                  & -0.965                  & 5.97                  & -1.16                  & 8.27                  & -1.12                   \\
                      & 2 & 3.87                  & -0.550                  & 6.19                  & -0.696                  & 8.39                  & -0.784                   \\
\hline
\end{tabular}
\caption{
  Polar $l=2$, 3 and 4 quasi-normal modes: dimensionless frequency $\omega_R r_0$ and dimensionless decay rate $\omega_I r_0$ vs dimensionless mass $M/r_0$ ($r_0=1$). 
 \label{table1}
  }
\end{table}

\begin{figure}
  \centering
  \includegraphics[angle =-90,width=0.48\textwidth]{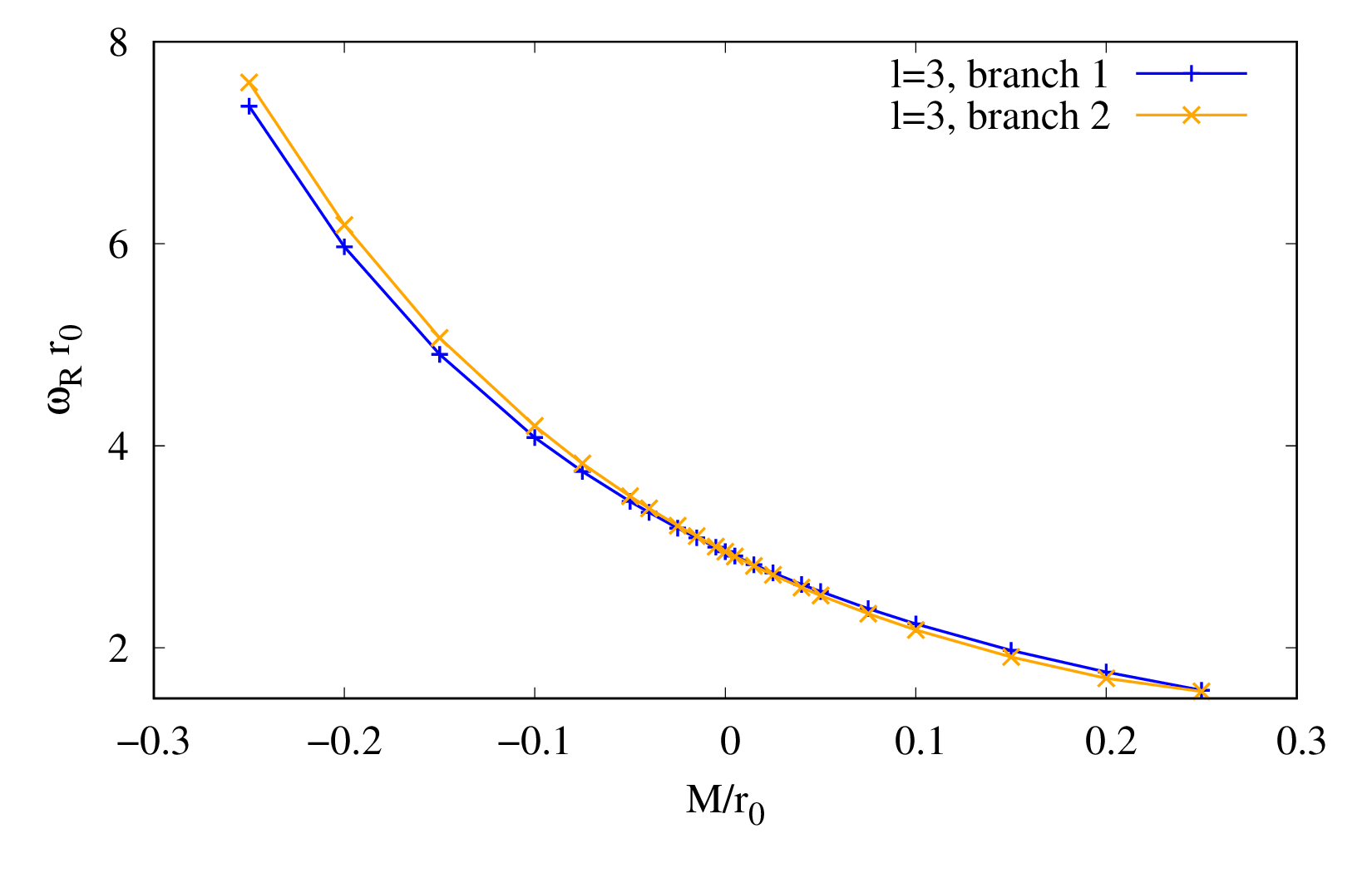}
  \includegraphics[angle =-90,width=0.48\textwidth]{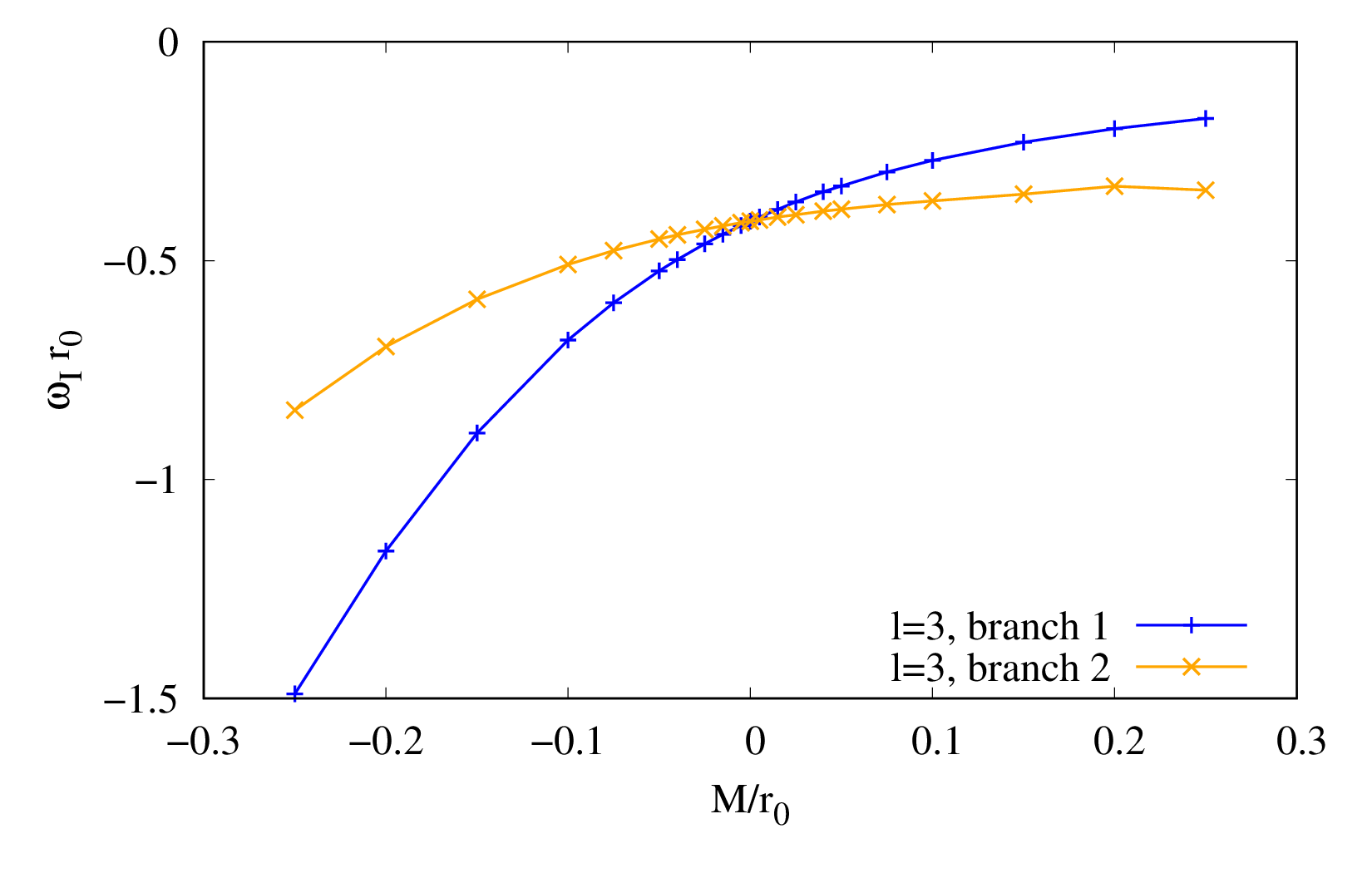}
  \caption{
  Polar $l=3$ quasi-normal modes: dimensionless frequency $\omega_R r_0$ (left) and dimensionless decay rate $\omega_I r_0$ (right) vs dimensionless mass $M/r_0$ ($r_0=1$). 
  }
  \label{Fig:polar_R_l3}
  \end{figure}

\begin{figure}
  \centering
  \includegraphics[angle =-90,width=0.48\textwidth]{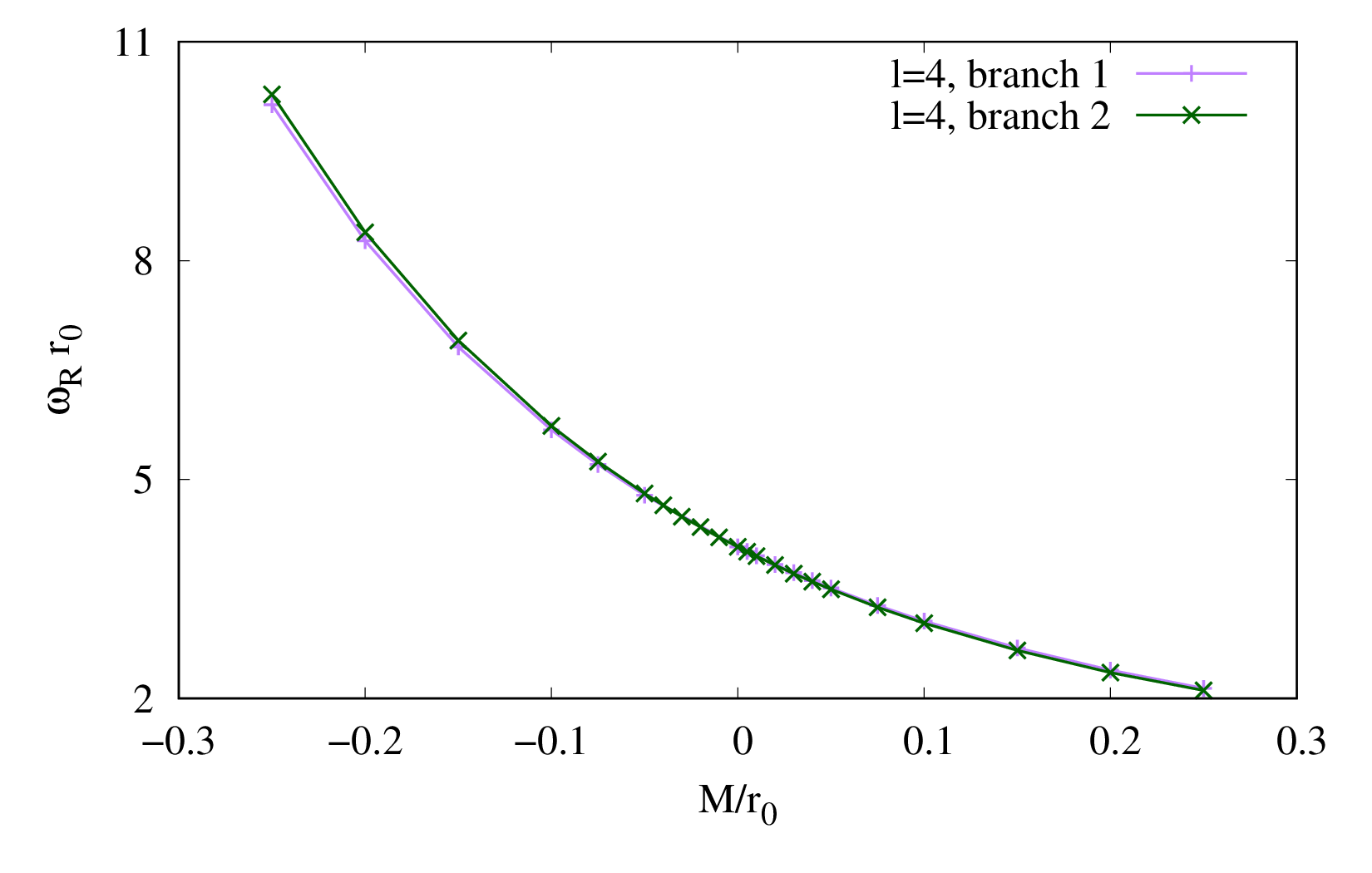}
  \includegraphics[angle =-90,width=0.48\textwidth]{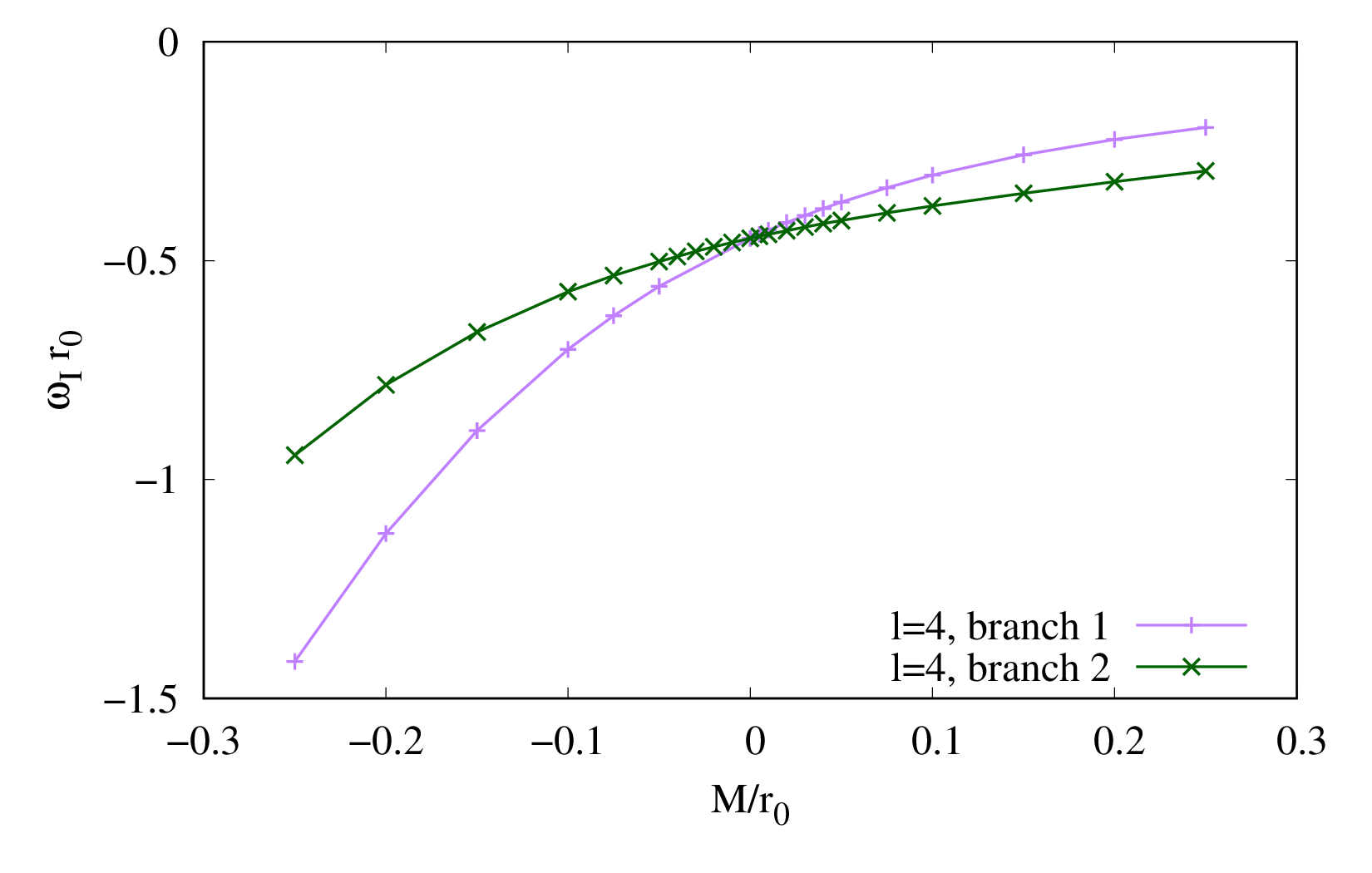}
  \caption{
  Polar $l=4$ quasi-normal modes: dimensionless frequency $\omega_R r_0$ (left) and dimensionless decay rate $\omega_I r_0$ (right) vs dimensionless mass $M/r_0$ ($r_0=1$). 
  }
  \label{Fig:polar_R_l4}
  \end{figure}

We exhibit the polar $l=3$ and $l=4$ modes in figure \ref{Fig:polar_R_l3} and \ref{Fig:polar_R_l4}, respectively.
Again we notice the crossing of the two branches in the massless case.
Away from the crossings the frequencies $\omega_R$ of both branches are getting closer for $l=3$ and even closer for $l=4$ than for the quadrupole.
The decay rates show a smoother behavior for the higher $l$ as compared to the quadrupole.
Table \ref{table1} shows again a selection of values for these quasi-normal modes.

\begin{figure}
  \centering
  \includegraphics[angle =-90,width=0.48\textwidth]{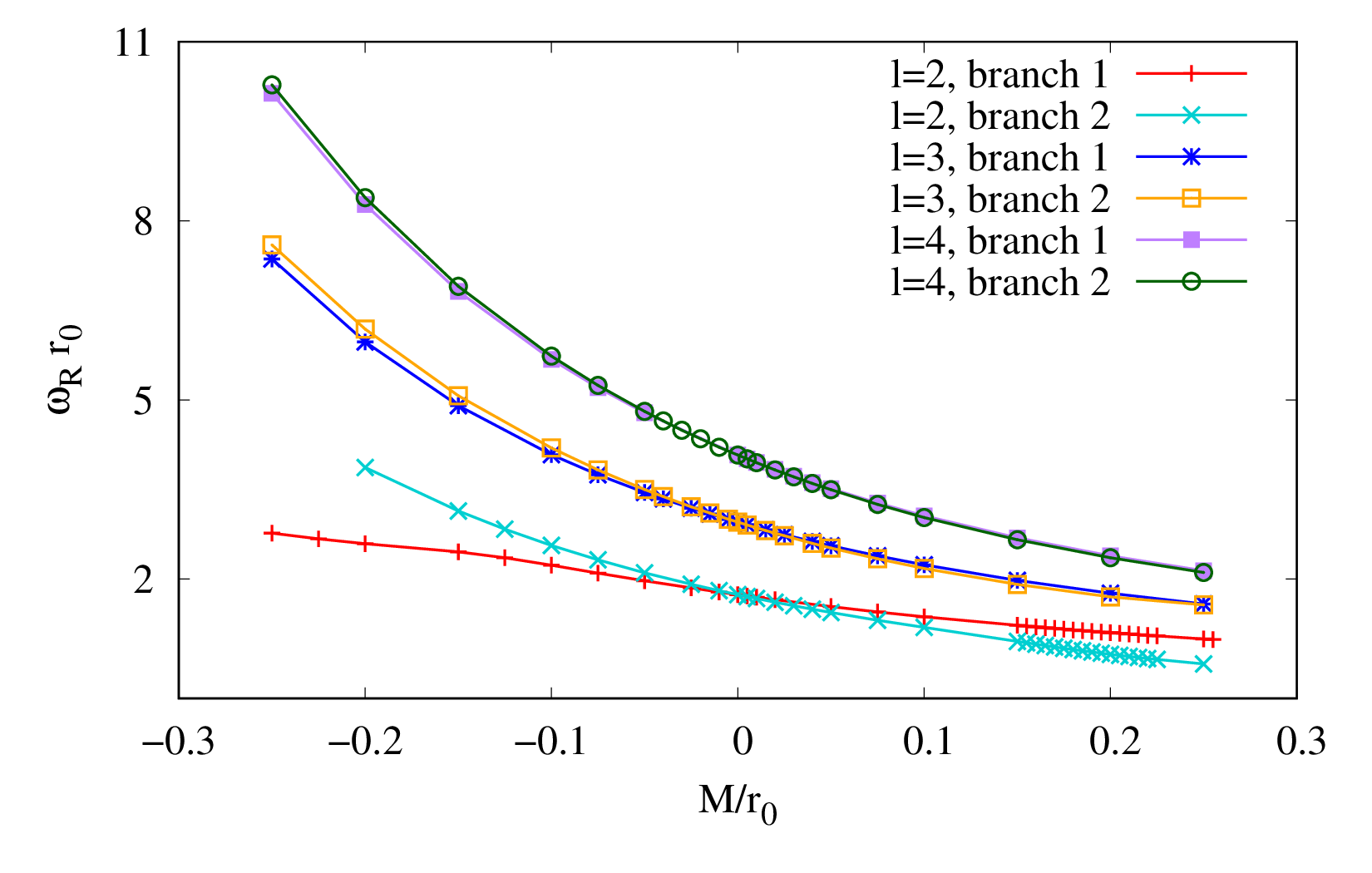}
  \includegraphics[angle =-90,width=0.48\textwidth]{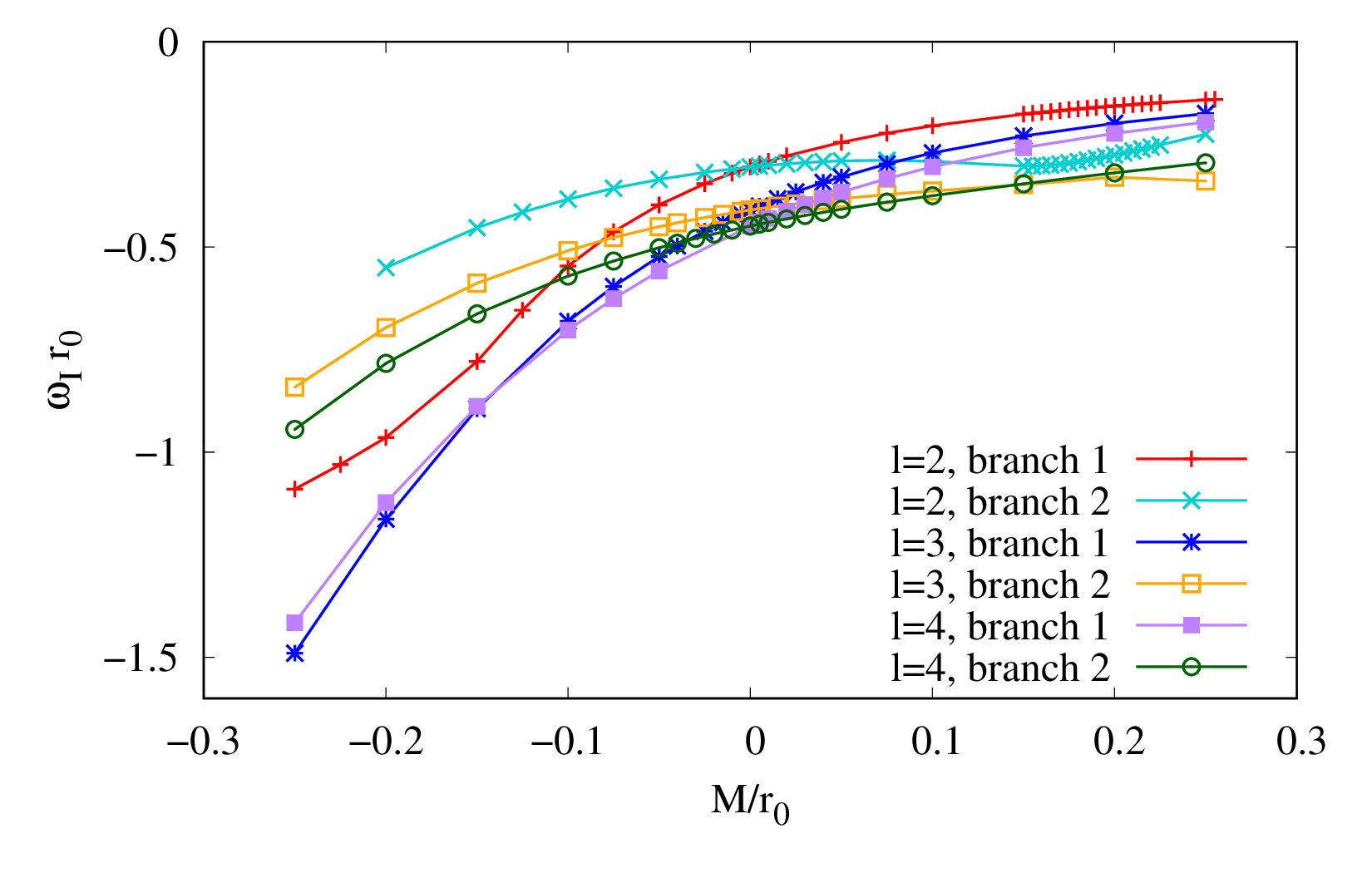}
  \caption{
  Comparison of polar $l=2$, 3 and 4 quasi-normal modes: dimensionless frequency $\omega_R r_0$ (left) and dimensionless decay rate $\omega_I r_0$ (right) vs dimensionless mass $M/r_0$ ($r_0=1$). 
  }
  \label{Fig:polar_R_lall}
  \end{figure}

We have collected the branches for $l=2$, 3 and 4 in figure \ref{Fig:polar_R_lall} to allow for a better comparison of the different $l$.
The left figure shows that the frequencies rise monotonically with increasing $l$.
From the right figure we note the overall trend on an increase of the decay rates with increasing $l$.
However, here the branches of the different $l$ intertwine.
For large positive masses, the decay rate of the branches 1 increases with $l$ and is smaller than decay rate of the branches 2, where the $l=4$ decay rate is smaller than the $l=3$ decay rate.
For large negative masses, the decay rate of the branches 2 increases with $l$ and is smaller than decay rate of the branches 1, where again the $l=4$ decay rate is smaller than the $l=3$ decay rate.
We note, that the decay rates of the branches 1 are very close for $l=3$ and $l=4$.

\begin{figure}
  \centering
  \includegraphics[angle =-90,width=0.48\textwidth]{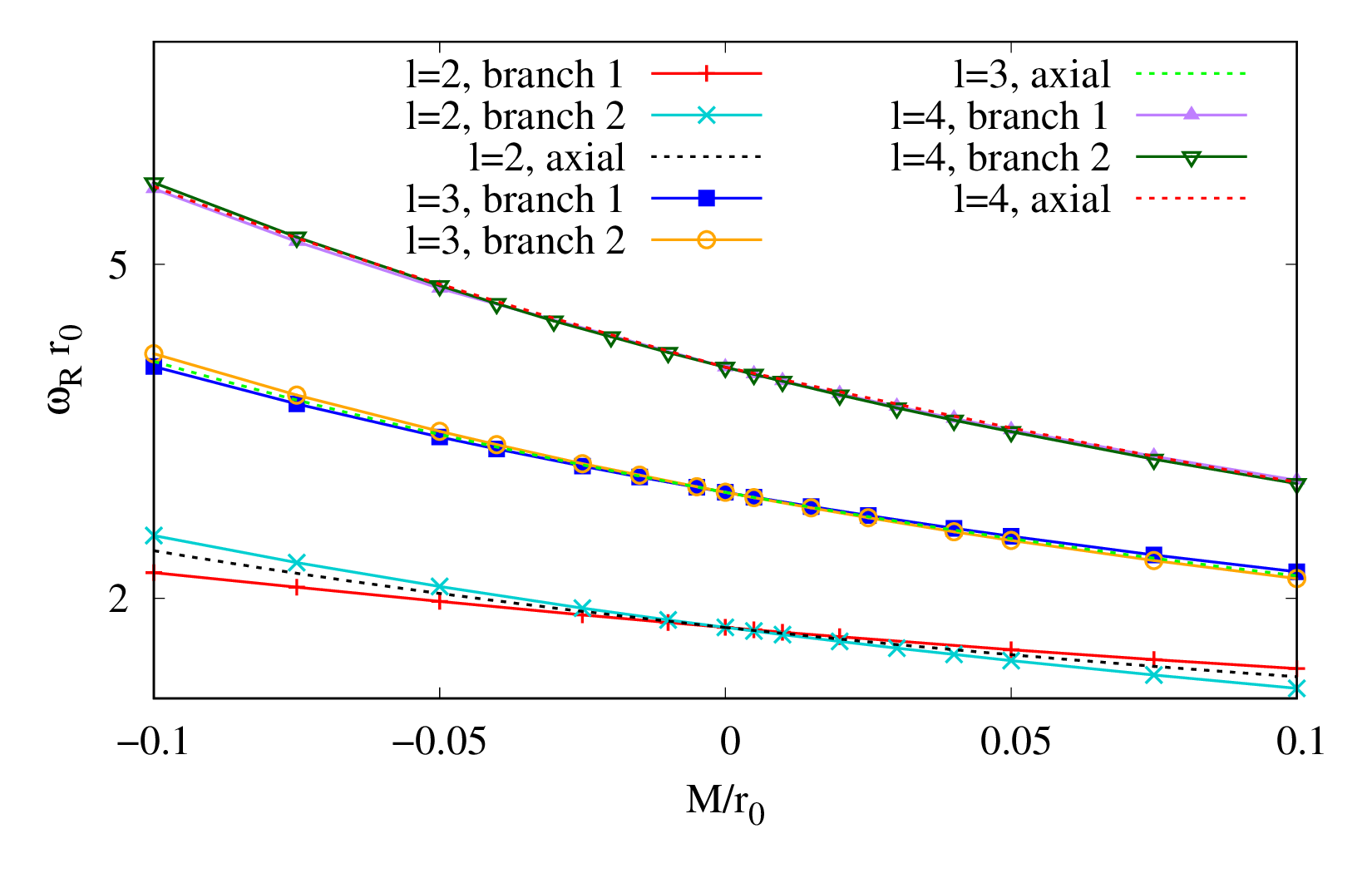}
  \includegraphics[angle =-90,width=0.48\textwidth]{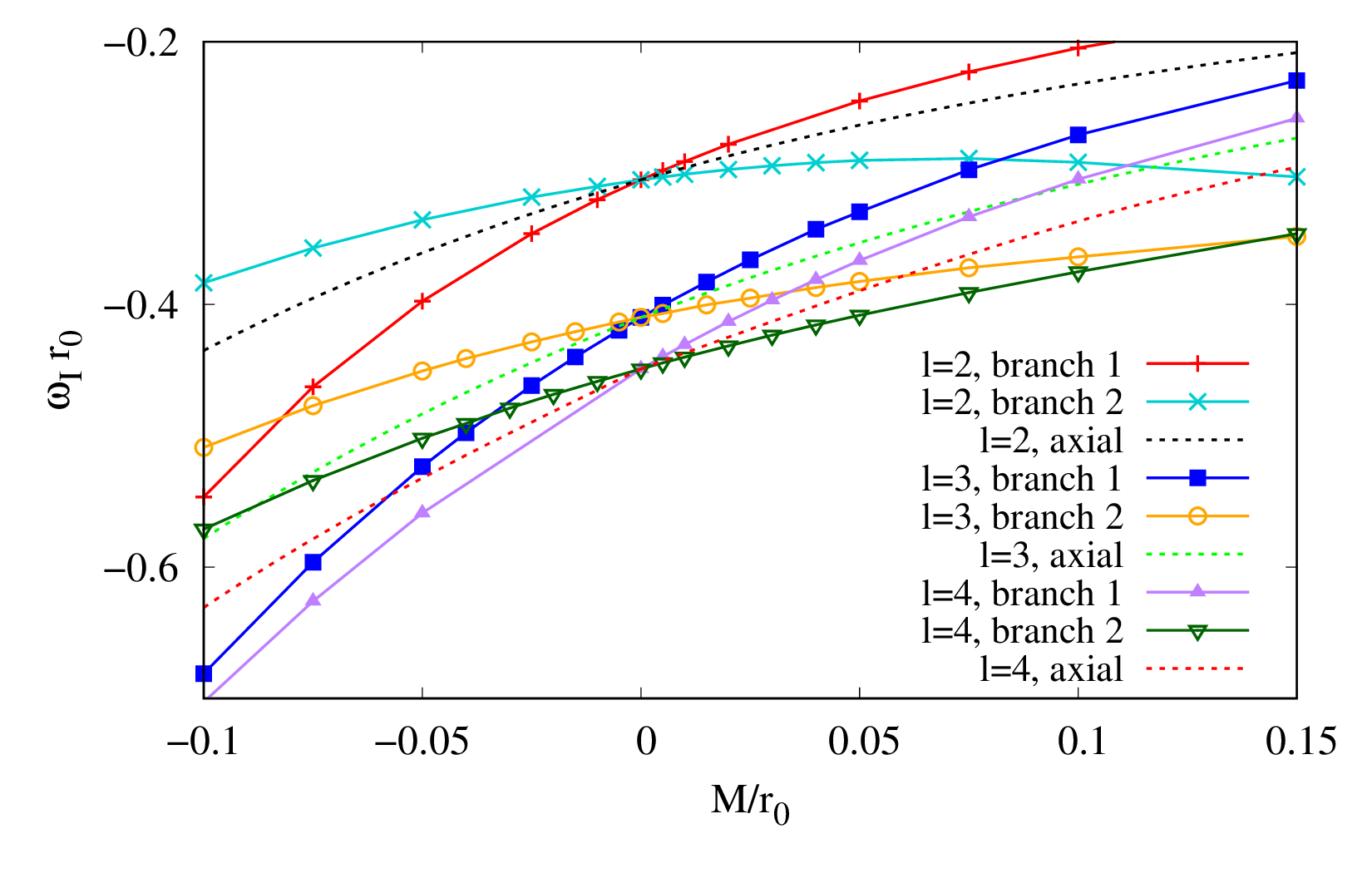}
  \caption{
  Comparison of axial and polar $l=2$, 3 and 4 quasi-normal modes: dimensionless frequency $\omega_R r_0$ (left) and dimensionless decay rate $\omega_I r_0$ (right) vs dimensionless mass $M/r_0$ ($r_0=1$). 
  }
  \label{Fig:polar_R_lall_ax}
  \end{figure}

Comparison with the $l \ge 2$ axial modes \cite{Blazquez-Salcedo:2018ipc} shows that the wormhole solutions do not possess isospectrality of their modes, as long as they possess a finite mass.
Only in the massless case the spectrum degenerates and all three fundamental branches for a given $l$ possess the same eigenvalue.
This is demonstrated in figure \ref{Fig:polar_R_lall_ax} for $l=2$, 3 and 4.

\subsubsection{Isospectrality
for $C=0$}

We now briefly comment on the mode degeneracy for the massless wormholes {from our numerical point of view}.
For $l>1$, we look for a linear combination of two independent solutions on the left side, $Z_i^-(I)$ and $Z_i^-(II)$, that smoothly matches a linear combination of two independent solutions on the right side, $Z_i^+(I)$ and $Z_i^+(II)$. Written explicitly, this implies that, at $r=r_c$
   \be 
   \det\bar{M}=0
   \ee
   where
    \be
    \bar{M}=\left[\begin{array}{cccc}
    u^{-}(I)&u^{-}(II)&u^{+}(I)&u^{+}(II)\\
    u'^{-}(I)&u'^{-}(II)&u'^{+}(I)&u'^{+}(II)\\
    H_1^{-}(I)&H_1^{-}(II)&H_1^{+}(I)&H_1^{+}(II)\\
    K^{-}(I)&K^{-}(II)&K^{+}(I)&K^{+}(II)
    \end{array}\right]_{r=r_c}
    \ee 
When $C\neq0$, this determinant becomes zero for two different values of $\omega$, one of them belonging to branch 1 and the other one belonging to branch 2
\footnote{Of course the determinant has more than two zeroes, corresponding to excited modes, with larger imaginary part of $\omega$. We here focus only on the fundamental modes.}.
At each of these zeroes of the determinant, there is a single linear combination of the perturbation functions that results in a smooth solution across all of the space-time (for each value of $\omega$, the Kernel of matrix $\bar M$ has dimension one). 
As we have seen in the previous subsection, as we decrease the value of $C$, the two branches of modes get closer and closer. 
What happens is that, in the limit when $C$ vanishes, the determinant possesses a single double-zero at one particular value of $\omega$. 
In this case it is possible to find, for this value of the eigenfrequency, two distinct linear combinations of the perturbation functions that result in two independent smooth solutions of the perturbation equation (the Kernel of matrix $\bar M$ has dimension two). 
In other words, the massless wormhole possesses a doubly-degenerate fundamental polar mode, as predicted in Section 3 in terms of the master equations.

\begin{figure}
  \centering
  \includegraphics[angle =-90,width=0.48\textwidth]{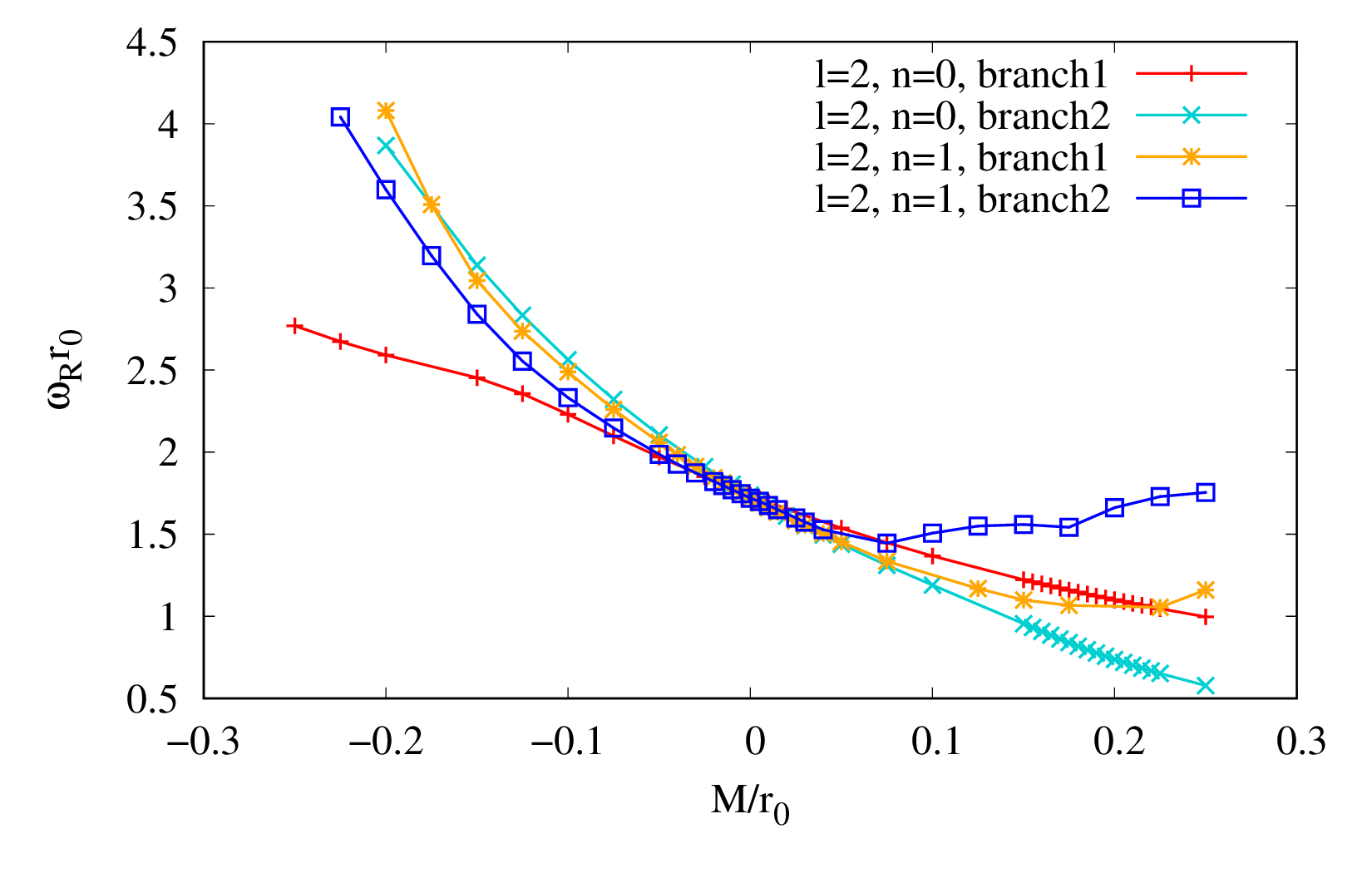}
  \includegraphics[angle =-90,width=0.48\textwidth]{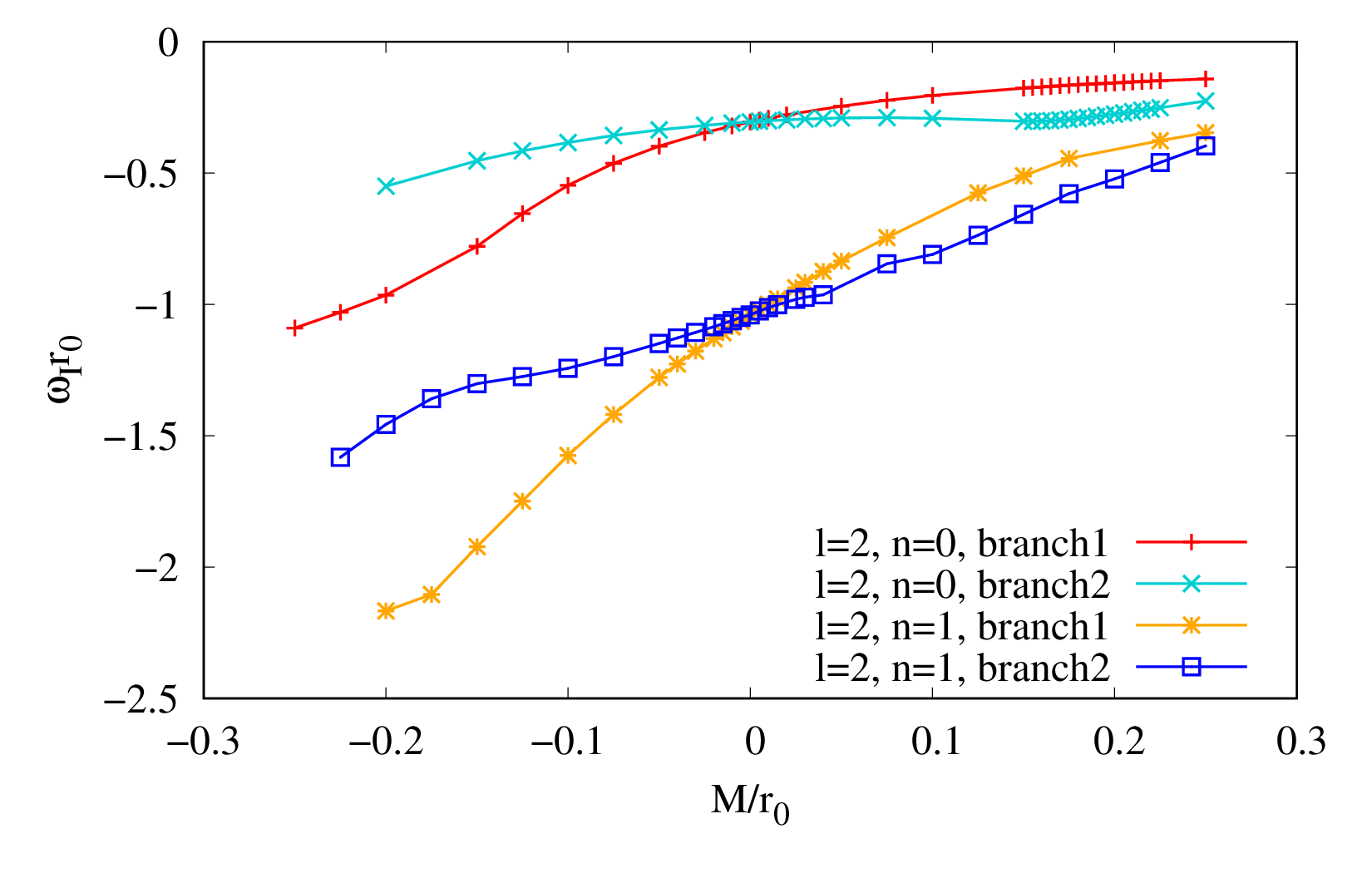}
  \caption{
  Fundamental ($n=0$) polar $l=2$ quasi-normal modes and their first overtone ($n=1$): dimensionless frequency $\omega_R r_0$ (left) and dimensionless decay rate $\omega_I r_0$ (right) vs dimensionless mass $M/r_0$ ($r_0=1$).
  }
  \label{Fig:polar_over}
  \end{figure}

In figure \ref{Fig:polar_over} we demonstrate the degeneracy of the polar modes in the massless case ($C=0$) for the first overtone ($n=1$) of $l=2$.
For comparison the figure also shows the corresponding fundamental ($n=0$) modes.

\subsection{$l=1$ and $l=0$}

To obtain the damped quasinormal modes for the $l=1$ and $l=0$ perturbations, we solve equations (\ref{eq_sc_l1}) and (\ref{eq_sc_l0}) respectively.
These equations can be cast into a second order Schr\"odinger-like equation, as discussed in Section 3
    \be   
    \frac{d^2Z}{dr^*}+\left(\omega^2-V_s(r)\right)Z=0 \,,
    \ee
where $V_s$ is the scalar perturbation potential for $l=0,1$ (see \cite{Blazquez-Salcedo:2018ipc}).
Assuming again, that the modes are purely outgoing at infinity $r^* \to \pm\infty$
    \be
    Z \simeq e^{\pm i\omega r^*},\  \  r^* \to \pm \infty \,,
     \ee 
we then solve the second order equation subject to these boundary conditions. 
The quasi-normal modes are obtained when the condition
\begin{equation}
 \frac{1}{Z^-} \frac{d Z^-}{d r^{*}}  \bigg|_{r=r_c} - \frac{1}{Z^+} \frac{d Z^+}{d r^{*}}  \bigg|_{r=r_c}  =0 \,
\end{equation}
is satisfied.

The polar $l=0$ and $l=1$ quasi-normal modes correspond to the scalar branches obtained previously \cite{Blazquez-Salcedo:2018ipc}.
Clearly, we obtain only one branch of fundamental modes for each of these values.
The frequencies of these modes increase with $l$, although their decay rates don't change much. 
For large masses the corresponding Schwarzschild scalar modes are approached.

For $l=0$ 
there is in addition an unstable mode 
\cite{Shinkai:2002gv,Gonzalez:2008wd,Gonzalez:2008xk,Cremona:2018wkj,Blazquez-Salcedo:2018ipc}.
In this case, the unstable mode can be obtained solving equation (\ref{eq_rad_pert}).
This unstable radial mode decreases in strength with increasing wormhole mass.
Thus the wormhole gets more stable as its mass increases, however, the wormhole always retains a radial instability as long as its mass is finite.

\section{Conclusions}

Here we have considered the polar modes of Ellis-Bronnikov wormholes.
We have obtained the sets of perturbation equations for general multipole number $l$. 
For $l\ge 2$ we have then solved numerically the corresponding  system of equations subject to purely outgoing boundary conditions at both radial infinities.
For $l=1$ and $l=0$ we have shown, that analogous to the case of the axial modes single master equations result.

The massless wormholes are special, however. 
Here isospectrality with a threefold degeneracy arises, since the set of polar equations can be reduced to the same master equation for both types of polar 
modes, which moreover agrees with the master equation for the axial modes, obtained previously \cite{Blazquez-Salcedo:2018ipc}.

For finite wormhole masses the spectrum of polar quasi-normal modes possesses two distinct branches for a given multipole number $l$.
For large wormhole masses these may {possibly be} associated with the scalar and the gravitational modes of a Schwarzschild black hole.
The modes exhibit an overall increase of the frequencies with the multipole number $l$.
Similarly, the decay rates show an overall increase with $l$.
For a given $l$, the two branches of polar modes also differ from the branch of axial modes, except for the degenerate massless case.

Since we now have access to the complete spectrum of quasi-normal modes of the static spherically symmetric Ellis-Bronnikov wormholes, we may next consider the inclusion of rotation.
This will be done perturbatively for small angular momenta analogous to \cite{Blazquez-Salcedo:2022eik}.
Of particular interest will, however, be the influence of rotation on the unstable radial mode of the Ellis-Bronnikov wormholes.
Could rotation have a stabilizing influence in four spacetime dimensions analogous to what has been observed before in five spacetime dimensions 
\cite{Dzhunushaliev:2013jja}?

\section*{Acknowledgements}

BA, JK and JLBS would like to gratefully acknowledge support by DAAD, the DFG Research Training Group 1620 \textit{Models of Gravity}, DFG project Ku612/18-1, FCT project PTDC/FIS-AST/3041/2020 and MICINN project PID2021-125617NB-I00 ``QuasiMode".
JLBS gratefully acknowledges support from Santander-UCM project PR44/21‐29910. XYC and DY are supported by the National Research Foundation of Korea (Grant No.: 2021R1C1C1008622, 2021R1A4A5031460).
We thank Fech Scen Khoo, Luis Manuel González-Romero and Francisco Navarro-Lérida for discussions.

\bibliographystyle{unsrt}

\begin{thebibliography}{99}

%

\bibitem{Einstein:1935tc}
A.~Einstein and N.~Rosen,
``The Particle Problem in the General Theory of Relativity,''
Phys. Rev. \textbf{48}, 73 (1935)

\bibitem{Ellis:1973yv} 
  H.~G.~Ellis,
  J.\ Math.\ Phys.\  {\bf 14}, 104 (1973)

\bibitem{Ellis:1979bh} 
  H.~G.~Ellis,
  Gen.\ Rel.\ Grav.\  {\bf 10}, 105 (1979).

\bibitem{Bronnikov:1973fh} 
  K.~A.~Bronnikov,
  Acta Phys.\ Polon.\ B {\bf 4}, 251 (1973).

\bibitem{Morris:1988cz}
M.~S.~Morris and K.~S.~Thorne,
``Wormholes in space-time and their use for interstellar travel: A tool for teaching general relativity,''
Am. J. Phys. \textbf{56}, 395-412 (1988)

\bibitem{Visser:1995cc} 
  M.~Visser,
  ``Lorentzian wormholes: From Einstein to Hawking,''
  Woodbury, USA: AIP (1995).


\bibitem{Alcubierre:2017pqm}
M.~Alcubierre and F.~S.~N.~Lobo,
Fundam. Theor. Phys. \textbf{189} (2017) 

\bibitem{Kanti:2011jz}
  P.~Kanti, B.~Kleihaus and J.~Kunz,
  Phys.\ Rev.\ Lett.\  {\bf 107}, 271101 (2011)

\bibitem{Kanti:2011yv}
  P.~Kanti, B.~Kleihaus and J.~Kunz,
  Phys.\ Rev.\ D {\bf 85}, 044007 (2012).

\bibitem{Antoniou:2019awm}
G.~Antoniou, A.~Bakopoulos, P.~Kanti, B.~Kleihaus and J.~Kunz,
Phys. Rev. D \textbf{101}, no.2, 024033 (2020)

\bibitem{Blazquez-Salcedo:2020czn}
J.~L.~Bl\'azquez-Salcedo, C.~Knoll and E.~Radu,
Phys. Rev. Lett. \textbf{126}, 101102 (2021)

\bibitem{Blazquez-Salcedo:2021udn}
J.~L.~Bl\'azquez-Salcedo, C.~Knoll and E.~Radu,
Eur. Phys. J. C \textbf{82}, 533 (2022)

\bibitem{Blazquez-Salcedo:2019uqq}
J.~L.~Bl\'azquez-Salcedo and C.~Knoll,
Eur. Phys. J. C \textbf{80}, no.2, 174 (2020)

\bibitem{Konoplya:2021hsm}
R.~A.~Konoplya and A.~Zhidenko,
Phys. Rev. Lett. \textbf{128}, 091104 (2022)

\bibitem{Barros:2018lca}
B.~J.~Barros and F.~S.~N.~Lobo,
Phys. Rev. D \textbf{98}, 044012 (2018)

\bibitem{Bouhmadi-Lopez:2021zwt}
M.~Bouhmadi-L\'opez, C.~Y.~Chen, X.~Y.~Chew, Y.~C.~Ong and D.~h.~Yeom,
JCAP \textbf{10} (2021), 059

\bibitem{Cramer:1994qj}
  J.~G.~Cramer, R.~L.~Forward, M.~S.~Morris, M.~Visser, G.~Benford and G.~A.~Landis,
  Phys.\ Rev.\ D {\bf 51}, 3117 (1995)

\bibitem{Safonova:2001vz}
  M.~Safonova, D.~F.~Torres and G.~E.~Romero,
  Phys.\ Rev.\ D {\bf 65}, 023001 (2002)

\bibitem{Perlick:2003vg}
  V.~Perlick,
  Phys.\ Rev.\ D {\bf 69}, 064017 (2004)

\bibitem{Nandi:2006ds}
  K.~K.~Nandi, Y.~Z.~Zhang and A.~V.~Zakharov,
  Phys.\ Rev.\ D {\bf 74}, 024020 (2006)

\bibitem{Abe:2010ap}
  F.~Abe,
  Astrophys.\ J.\  {\bf 725}, 787 (2010)

\bibitem{Toki:2011zu}
  Y.~Toki, T.~Kitamura, H.~Asada and F.~Abe,
  Astrophys.\ J.\  {\bf 740}, 121 (2011)

\bibitem{Nakajima:2012pu}
  K.~Nakajima and H.~Asada,
  Phys.\ Rev.\ D {\bf 85}, 107501 (2012)

\bibitem{Tsukamoto:2012xs}
  N.~Tsukamoto, T.~Harada and K.~Yajima,
  Phys.\ Rev.\ D {\bf 86}, 104062 (2012)

\bibitem{Kuhfittig:2013hva}
  P.~K.~F.~Kuhfittig,
  Eur. Phys.\ J. \ C {\bf 74},  2818 (2014)
 
\bibitem{Bambi:2013nla}
  C.~Bambi,
  Phys.\ Rev.\ D {\bf 87}, 107501 (2013)

\bibitem{Takahashi:2013jqa}
  R.~Takahashi and H.~Asada,
  Astrophys.\ J.\  {\bf 768}, L16 (2013)

\bibitem{Tsukamoto:2016zdu}
  N.~Tsukamoto and T.~Harada,
  Phys.\ Rev.\ D {\bf 95},   024030 (2017)

\bibitem{Nedkova:2013msa}
  P.~G.~Nedkova, V.~K.~Tinchev and S.~S.~Yazadjiev,
  Phys.\ Rev.\ D {\bf 88},   124019 (2013)

\bibitem{Ohgami:2015nra}
  T.~Ohgami and N.~Sakai,
  Phys.\ Rev.\ D {\bf 91}, 124020 (2015)

\bibitem{Shaikh:2018kfv}
  R.~Shaikh,
  Phys.\ Rev.\ D {\bf 98},  024044 (2018)

\bibitem{Gyulchev:2018fmd}
  G.~Gyulchev, P.~Nedkova, V.~Tinchev and S.~Yazadjiev,
  Eur.\ Phys.\ J.\ C {\bf 78} ,  544 (2018).

\bibitem{Guerrero:2022qkh}
M.~Guerrero, G.~J.~Olmo, D.~Rubiera-Garcia and D.~G\'omez S\'aez-Chill\'on,
Phys. Rev. D \textbf{105}, no.8, 084057 (2022)

\bibitem{Harko:2008vy}
  T.~Harko, Z.~Kovacs and F.~S.~N.~Lobo,
  Phys.\ Rev.\ D {\bf 78}, 084005 (2008)

\bibitem{Harko:2009xf}
  T.~Harko, Z.~Kovacs and F.~S.~N.~Lobo,
  Phys.\ Rev.\ D {\bf 79}, 064001 (2009)

\bibitem{Bambi:2013jda}
  C.~Bambi,
  Phys.\ Rev.\ D {\bf 87}, 084039 (2013)

\bibitem{Zhou:2016koy}
  M.~Zhou, A.~Cardenas-Avendano, C.~Bambi, B.~Kleihaus and J.~Kunz,
  Phys.\ Rev.\ D {\bf 94}, 024036 (2016)

\bibitem{Lamy:2018zvj}
  F.~Lamy, E.~Gourgoulhon, T.~Paumard and F.~H.~Vincent,
  Class.\ Quant.\ Grav.\  {\bf 35},   115009 (2018)

\bibitem{Deligianni:2021ecz}
E.~Deligianni, J.~Kunz, P.~Nedkova, S.~Yazadjiev and R.~Zheleva,
Phys. Rev. D \textbf{104},  024048 (2021)

\bibitem{Deligianni:2021hwt}
E.~Deligianni, B.~Kleihaus, J.~Kunz, P.~Nedkova and S.~Yazadjiev,
Phys. Rev. D \textbf{104},  064043 (2021)

\bibitem{Azad:2020ajs}
B.~Azad, F.~Loran and A.~Mostafazadeh,
``Transmission of low-energy scalar waves through a traversable wormhole,''
Eur. Phys. J. C \textbf{80}, no.12, 1097 (2020)

\bibitem{LIGOScientific:2016aoc}
B.~P.~Abbott \textit{et al.} [LIGO Scientific and Virgo],
Phys. Rev. Lett. \textbf{116}, no.6, 061102 (2016).
doi:10.1103/PhysRevLett.116.061102.

\bibitem{LIGOScientific:2017vwq}
B.~P.~Abbott \textit{et al.} [LIGO Scientific and Virgo],
Phys. Rev. Lett. \textbf{119} , no.16, 161101 (2017)
doi:10.1103/PhysRevLett.119.161101.

\bibitem{LIGOScientific:2017ync}
B.~P.~Abbott \textit{et al.} 
Astrophys. J. Lett. \textbf{848}, no.2, L12 (2017).

\bibitem{Kokkotas:1999bd}
K.~D.~Kokkotas and B.~G.~Schmidt,
Living Rev. Rel. \textbf{2}, 2 (1999)

\bibitem{Berti:2009kk}
E.~Berti, V.~Cardoso and A.~O.~Starinets,
Class. Quant. Grav. \textbf{26}, 163001 (2009)

\bibitem{Konoplya:2011qq}
R.~A.~Konoplya and A.~Zhidenko,
Rev. Mod. Phys. \textbf{83}, 793 (2011)



\bibitem{Konoplya:2005et}
  R.~A.~Konoplya and C.~Molina,
  Phys.\ Rev.\ D {\bf 71}, 124009 (2005)

\bibitem{Kim:2008zzj}
  S.~W.~Kim,
  Prog.\ Theor.\ Phys.\ Suppl.\  {\bf 172}, 21 (2008)

\bibitem{Konoplya:2010kv}
  R.~A.~Konoplya and A.~Zhidenko,
  Phys.\ Rev.\ D {\bf 81}, 124036 (2010)

\bibitem{Konoplya:2016hmd}
  R.~A.~Konoplya and A.~Zhidenko,
  JCAP {\bf 1612}, 043 (2016)
  
\bibitem{Volkel:2018hwb}
  S.~H.~V\"olkel and K.~D.~Kokkotas,
  Class.\ Quant.\ Grav.\  {\bf 35}, 105018 (2018)

\bibitem{Aneesh:2018hlp}
S.~Aneesh, S.~Bose and S.~Kar,
Phys. Rev. D \textbf{97}, 124004 (2018)

\bibitem{Konoplya:2018ala}
R.~A.~Konoplya,
Phys. Lett. B \textbf{784}, 43 (2018)

\bibitem{Blazquez-Salcedo:2018ipc}
J.~L.~Bl\'azquez-Salcedo, X.~Y.~Chew and J.~Kunz,
Phys. Rev. D \textbf{98}, 044035 (2018)

\bibitem{Konoplya:2019hml}
R.~A.~Konoplya, A.~F.~Zinhailo and Z.~Stuchl\'\i{}k,
Phys. Rev. D \textbf{99}, 124042 (2019)

\bibitem{Churilova:2019qph}
M.~S.~Churilova, R.~A.~Konoplya and A.~Zhidenko,
Phys. Lett. B \textbf{802}, 135207 (2020)

\bibitem{Jusufi:2020mmy}
K.~Jusufi,
Gen. Rel. Grav. \textbf{53}, 87 (2021)

\bibitem{Gonzalez:2022ote}
P.~A.~Gonz\'alez, E.~Papantonopoulos, \'A.~Rinc\'on and Y.~V\'asquez,
Phys. Rev. D \textbf{106}, 024050 (2022)

\bibitem{Chandrasekhar:1985kt}
S. Chandrasekhar,
\emph{The Mathematical Theory of Black Holes}, 
(Clarendon Press, Oxford, 1998)

\bibitem{Regge:1957td}
T.~Regge and J.~A.~Wheeler,
``Stability of a Schwarzschild singularity,''
Phys. Rev. \textbf{108}, 1063 (1957)

\bibitem{Ascher:1979iha}
U.~Ascher, J.~Christiansen and R.~D.~Russell,
Math. Comput. \textbf{33}, 659 (1979)

\bibitem{Blazquez-Salcedo:2016enn}
J.~L.~Bl\'azquez-Salcedo, C.~F.~B.~Macedo, V.~Cardoso, V.~Ferrari, L.~Gualtieri, F.~S.~Khoo, J.~Kunz and P.~Pani,
Phys. Rev. D \textbf{94}, 104024 (2016)

\bibitem{Shinkai:2002gv}
  H.~a.~Shinkai and S.~A.~Hayward,
  Phys.\ Rev.\ D {\bf 66}, 044005 (2002)

\bibitem{Gonzalez:2008wd}
  J.~A.~Gonzalez, F.~S.~Guzman and O.~Sarbach,
  Class.\ Quant.\ Grav.\  {\bf 26}, 015010 (2009)

\bibitem{Gonzalez:2008xk}
  J.~A.~Gonzalez, F.~S.~Guzman and O.~Sarbach,
  Class.\ Quant.\ Grav.\  {\bf 26},  015011 (2009)


\bibitem{Cremona:2018wkj}
F.~Cremona, F.~Pirotta and L.~Pizzocchero,
Gen. Rel. Grav. \textbf{51}, 19 (2019)

\bibitem{Blazquez-Salcedo:2022eik}
J.~L.~Bl\'azquez-Salcedo and F.~S.~Khoo,
[arXiv:2212.00054 [gr-qc]].

\bibitem{Dzhunushaliev:2013jja}
V.~Dzhunushaliev, V.~Folomeev, B.~Kleihaus, J.~Kunz and E.~Radu,
Phys. Rev. D \textbf{88}, 124028 (2013)



\end{thebibliography}

\end{document}